\documentclass{article}

\usepackage{arxiv}
\usepackage{geometry}
\usepackage[utf8]{inputenc} 
\usepackage[T1]{fontenc}    
\usepackage{hyperref}       
\usepackage{url}            
\usepackage{booktabs}       
\usepackage{amsfonts}       
\usepackage{nicefrac}       
\usepackage{microtype}      
\usepackage{graphicx}
\usepackage{amsmath}
\usepackage{siunitx}

\usepackage[ruled,vlined]{algorithm2e}
\SetKwRepeat{Struct}{struct \{}{\}}%
\newcommand{\Float}{\KwSty{float}}
\newcommand{\String}{\KwSty{string}}
\newcommand{\Int}{\KwSty{int}}
\newcommand{\Enum}{\KwSty{enum}}
\newcommand{\Vect}{\KwSty{vect}}
\newcommand{\Quat}{\KwSty{quat}}
\usepackage{listings}
\usepackage{setspace}

\newcommand{\eg}{e.\,g.}
\newcommand{\ie}{i.\,e.}
\newcommand{\vect}[1]{\mathbf{#1}}

\title{\textit{Modeling in the Time of COVID-19:}\\Statistical and Rule-based Mesoscale Models}

\usepackage{authblk}
\author[1]{Ngan~Nguyen}
\author[1]{Ond\v{r}ej~Strnad}
\author[2,4]{Tobias~Klein}
\author[1]{Deng~Luo}
\author[1]{Ruwayda~Alharbi}
\author[1]{Peter~Wonka}
\author[3]{Martina Maritan}
\author[2,4]{Peter Mindek}
\author[3]{Ludovic Autin}
\author[3]{David~S.~Goodsell}
\author[1]{Ivan~Viola}
\affil[1]{King Abdullah University of Science and Technology (KAUST), Saudi Arabia. E-mails: \{ngan.nquyen $\vert$ ondrej.strnad $\vert$ deng.luo $\vert$ ruwayda.alharbi $\vert$ peter.wonka $\vert$ ivan.viola~\}@kaust.edu.sa.
	\\N. Nguyen and O. Strnad are co-first authors.}
\affil[2]{TU Wien, Austria. E-mails: \{tklein $\vert$ mindek\}@cg.tuwien.ac.at.}
\affil[3]{Scripps Research Institute, US. E-mail: \{mmaritan $\vert$ autin $\vert$ goodsell\}@scripps.edu.}
\affil[4]{Nanographics GmbH}
\date{}                     
\newcommand{\annot}[1]{
}

\begin{document}
\maketitle

\begin{abstract}
	We present a new technique for rapid modeling and construction of scientifically accurate mesoscale biological models. Resulting 3D models are based on few 2D microscopy scans and the latest knowledge about the biological entity represented as a set of geometric relationships.
	Our new technique is based on statistical and rule-based modeling approaches that are rapid to author, fast to construct, and easy to revise. From a few 2D microscopy scans, we learn statistical properties of various structural aspects, such as the outer membrane shape, spatial properties and distribution characteristics of the macromolecular elements on the membrane. This information is utilized in 3D model construction. Once all imaging evidence is incorporated in the model, additional information can be incorporated by interactively defining rules that spatially characterize the rest of the biological entity, such as mutual interactions among macromolecules, their distances and orientations to other structures. These rules are defined through an intuitive 3D interactive visualization and modeling feedback loop.
	We demonstrate the utility of our approach on a use case of the modeling procedure of the SARS-CoV-2 virus particle ultrastructure. Its first complete atomistic model, which we present here, can steer biological research to new promising directions in fighting spread of the virus.
\end{abstract} 


\section{Introduction}
\begin{figure*}
	\centering
	\includegraphics[width=0.33\linewidth]{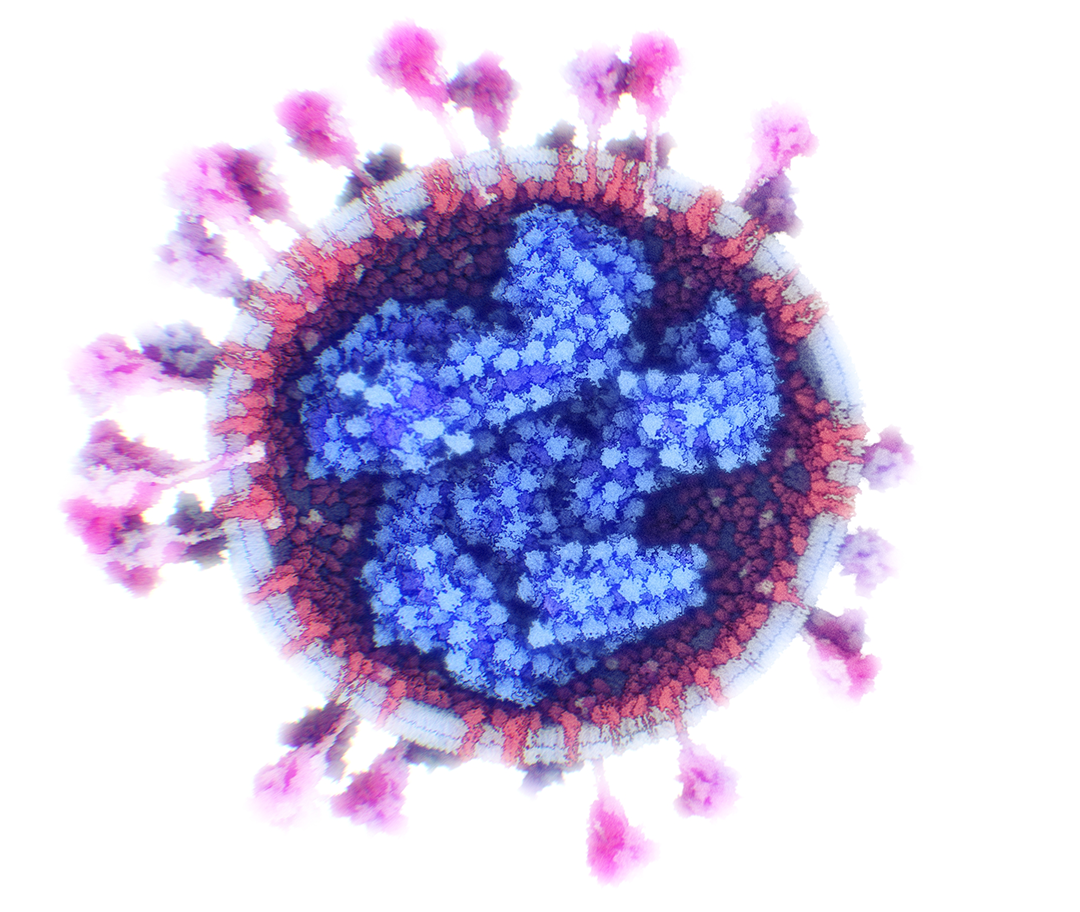}%
	\includegraphics[width=0.33\linewidth]{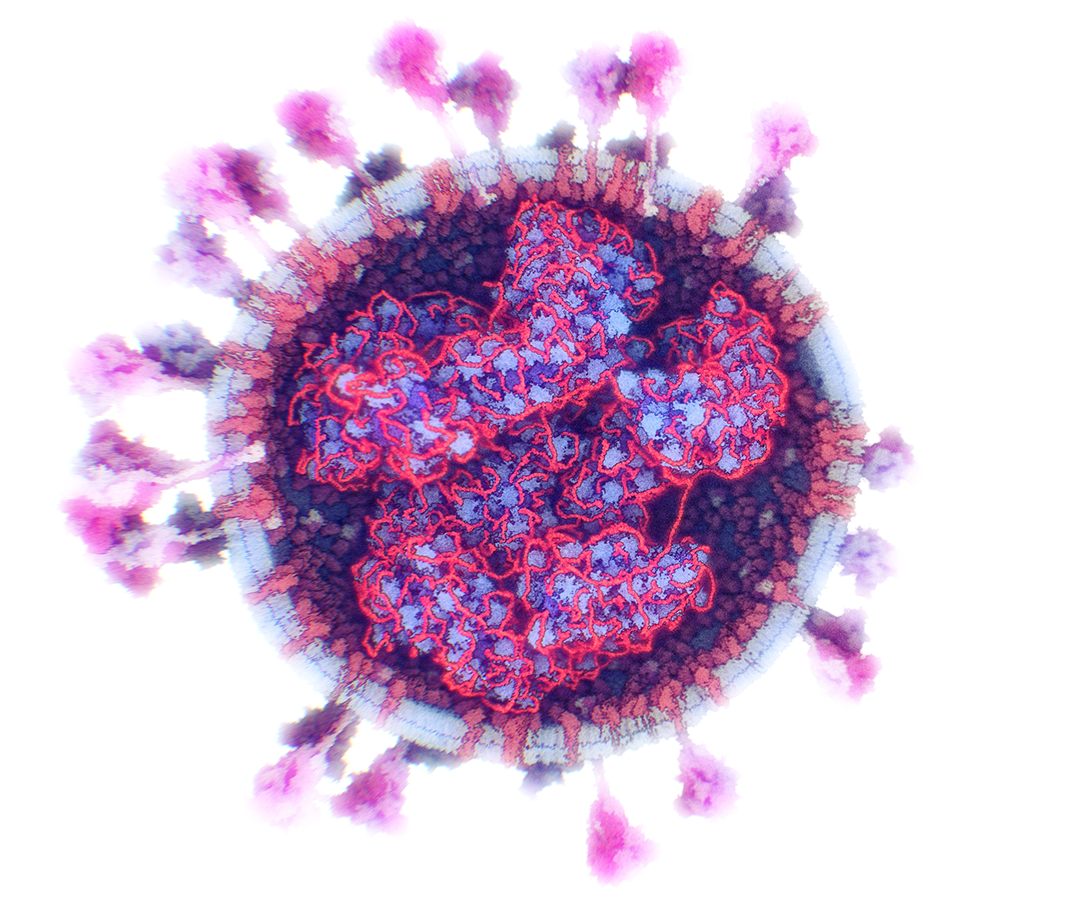}%
	\includegraphics[width=0.33\linewidth]{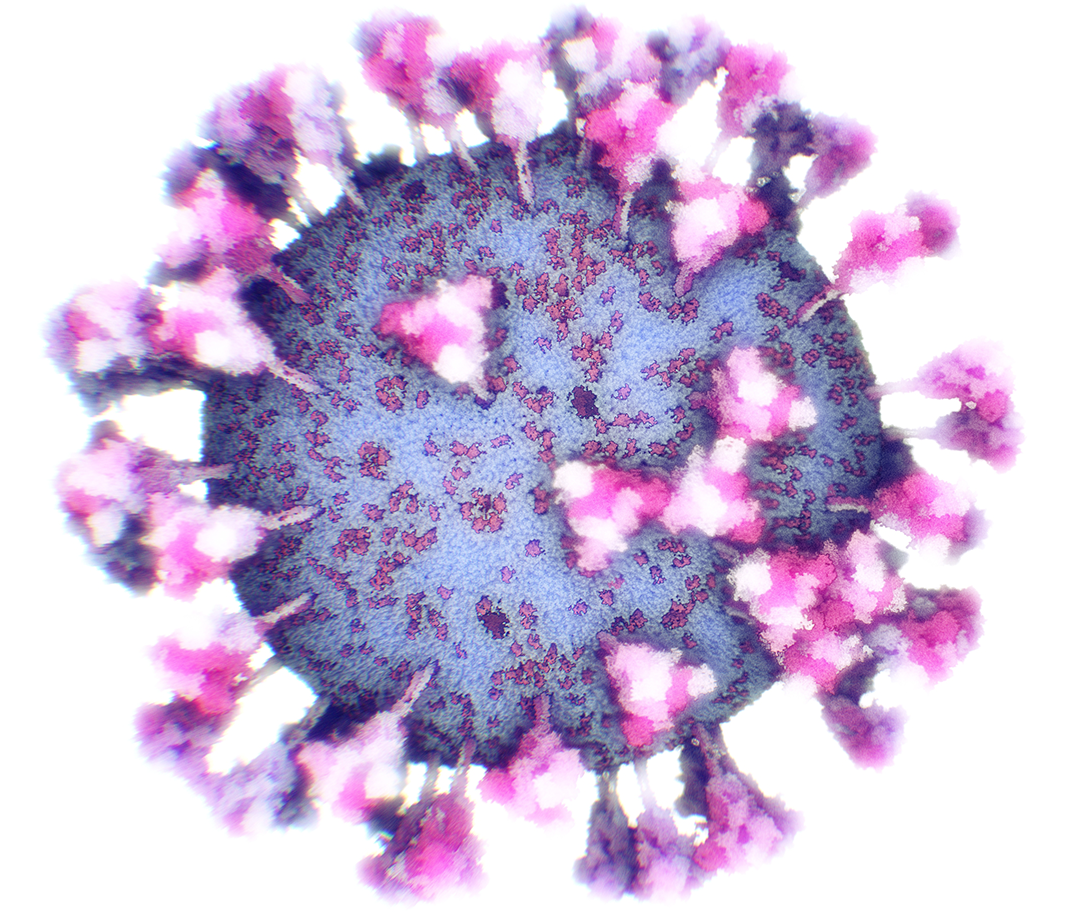}%
	\caption{The first complete ultrastructure of a SARS-CoV-2 particle created with our modeling technique. The membrane shape and distribution of spike proteins is learned from microscopy image data and the internal assembly is a result of an interactive 3D rule specification approach. Left: internal structure of the virus - the rope. Middle: RNA bound to the rope. Right: Overview of the model.}
	\label{fig:teaser}
\end{figure*}

Living organisms on Earth share a common complex, hierarchical structure. At the lowest level of the hierarchy, biomolecules such as proteins and DNA perform all of the basic nanoscale tasks of information management, energy transformation, directed motion, etc. These biomolecules are assembled into cells, the basic units of life. Cells typically are surrounded by a lipid bilayer membrane, which encloses several thousand different types of biomolecules that choreograph the processes of finding resources, responding to environmental changes, and ultimately growing and reproducing. Most familiar organisms, such as plants and animals, add an additional level to the hierarchy, with multiple cells cooperating to form large, multi-cellular organisms.

Viruses are pared-down versions of living organisms, with just enough of this hierarchical structure to perform a targeted task: to get inside a cell and force it to create more copies of itself. Viruses are typically comprised of some form of nucleic acid (RNA or DNA) that encodes the genome, and a small collection of proteins that are encoded in this genome, that form the molecular mechanism for finding cells and infecting them. Some viruses also include a surrounding envelope composed of a lipid bilayer membrane that is acquired as the virus buds from an infected cell.

Effective computational methods are available for modeling and visualizing the biomolecular components of cells and viruses. Atomic structures of over a hundred thousand biomolecules are available at the Protein Data Bank (wwpdb.org), and decades of research and development have generated a comprehensive toolbox of simulation, structure prediction, modeling, and visualization tools to utilize and extend this data~\cite{ODonoghue2010,Pirhadi2016}. However, modeling and visualization of the full hierarchical structure of living organisms---from atoms to cells---is a field still in its infancy, limited largely by the size and complexity of the hierarchy and its many interacting parts.

Modeling and visualization of the cellular mesoscale---the scale level bridging the nanoscale of atoms and molecules with the microscale of cells---is necessarily an integrative process, since there are no existing experimental methods for directly observing the mesoscale structure of cells~\cite{Goodsell2020}. Mesoscale studies integrate information from microscopy, structural biology, and bioinformatics to generate representative models consistent with the current state of knowledge. Challenges that are currently limiting the integrative modeling pipeline include (a) finding and curating disparate sources of data, and (b) intuitive 3D construction and visualization of models of this size and complexity within a reasonable user and computational effort. This latter challenge is addressed in this paper.

The central idea behind our rapid modeling approach for mesoscale models is to take advantage of the hierarchical structure of living systems. We model structural characteristics on a small representative collection of structural elements, which are then assembled into the entire cellular or viral system through a set of learned rules that guide placement and interaction of the component elements. These rules are specified directly through 3D interactive modeling, instead of indirectly through some rule-definition syntax. In this way, we can reduce the burden on the users, provide them with an intuitive modeling interface, and automatically generate instances of the full model comprised of a huge number of interactive component elements. In case the model needs to be further fine-tuned or new information needs to be incorporated, the construction rules are revised in 3D and new models are generated that incorporate the latest revision. If structural evidence is available in form of EM images, our system learns basic structural properties from images by few inputs from the user.

We demonstrate the rapid modeling method for integrating data from electron microscopy with structural information for the novel coronavirus SARS-CoV-2. Generated models can be used for exploring the diversity of structure and analyzing the detailed arrangement of spike glycoproteins on its surface.

\section{Related Work}
\annot{molecular modeling}Modeling geometric representations of molecules has been a driving scientific visualization and computer graphics research for several decades. Since late seventies Richards has developed a geometric representation of molecular surface that characterized their area~\cite{Richards1977}, which as a concept was popularized by Connolly \cite{Connolly1983}. Over the years many geometric construction algorithms for molecular surfaces have been developed, notably Reduced Surface~\cite{Sanner1996}, blobby objects~\cite{Blinn1982,Parulek2013},  or $\alpha$-shapes~\cite{Edelsbrunner1994} to name a few. These algorithms are typically well parallelizable on multiprocessor systems~\cite{Varshney1994} or on modern GPUs~\cite{Krone2009,Hermosilla2017,Chavent2011}, and nowadays scale up to interactive rates of huge atomistic models thanks to \eg view-guided rendering strategies~\cite{Bruckner2019}. Simplified representations, such as van-der-Waals space filling molecular models can be interactively constructed and visualized to represent up to billion-sized atomistic scenes ~\cite{Lindow2012,Falk2013,Klein2018} making use of various acceleration strategies, such as the procedural impostors~\cite{Tarini2006}, adaptive level-of-detail tesselation~\cite{LeMuzic2014,LeMuzic2015}, or hybrid particle-volumetric representation~\cite{Schatz2016}.

\annot{lipids and fibers}Geometric representations of molecular structures described above has been mostly concerned with modeling protein macromolecules. Recently, dedicated approaches for modeling large lipid membranes have been developed~\cite{Durrant2014,Bhatia2019}.
Another special type are fibrous macromolecules such as the 3D genome. Halladjian et al. presented an approach to construct and visualize a multi-scale model of interphase chromosomes~\cite{Halladjian2020}.
A typical approach to model the backbone of linear polymers like RNA or DNA procedurally is to concatenate building blocks with processes like a random walk. A random walk produces a sequence of points where the location of each generated point is dependent on its predecessor. While this process leads to plausible models and is able to incorporate measured characteristics like the stiffness, it is hard to control and guide to specific points. Klein et al.~\cite{Klein2019} propose a parallel algorithm for constructing a 3D genome sequence, which builds on the midpoint-displacement concept. We utilize this approach in calculating the path for genetic macromolecules.

\annot{rapid modeling}Above methods model molecular geometry based on some underlying well-defined structure. Technique presented in this paper are primarily concerned with interactive 3D modeling of the molecular assembly, which relates to methodologies developed in graphics research. A particular category is rapid 3D modeling, which can be characterized as a process where the author specifies the desired 3D model through a minimal amount of user interactions. The algorithm or learned statistical model constructs then the geometric model by preserving user-defined constraints. There are two dominant strategies for achieving rapid modeling.

\annot{sketch-based modeling}One methodology, known as sketch-based modeling, allows the user to specify certain geometric details directly in the scene. A good example for sketch-based modeling is the Teddy system presented by Igarashi et al.~\cite{Igarashi1999}. Here the user only specifies a 2D contour of an object and a 3D geometry is generated using the contour inflation approach~\cite{Williams1991}, which we also utilize. An interesting recent trend is to control a deep learning model using sketches, e.g. for modeling terrains~\cite{Guerin2017}, faces~\cite{Portenier2018}, or buildings~\cite{Nishida2016}.
Utilization in the sciences can be exemplified through modeling advanced geological concepts and phenomena~\cite{Lidal2013,Natali2014} or for creating quick molecular landscapes for communicating to peers or a broader audience~\cite{Gardner2018}. These works result in approximate \emph{sketches} of complex scientific scenarios and make use of rules that are algorithmically defined within the system. Our approach also features sketching components, however, the proposed approach is initially rule-free and all rules are specified by the user or determined from imaging data. Moreover, these rules can be defined at a wide range of modeling precision. Our rule design space is open: many simple rules are  generated first, then can be combined in a construction of more complex structural elements, and obsolete concepts can be revised into new rules and effortlessly reapplied. The established rules can be stored as structural templates that can be shared among users.

\annot{procedural modeling}The second approach is known as procedural modeling and its basic idea is that the geometric structure is defined indirectly by specifying rules and parameters of these rules. The rules are then used when executing  procedural construction of the 3D scene geometry, often without any direct geometric input from the user. The \emph{rapid} modeling aspect is achieved through quick setting of few parameters that can serve as sufficient input for massively large scenes.
Procedural modeling has a long tradition in computer graphics~\cite{Ebert2002}. It is frequently used for modeling large environments that \emph{look} plausible. Examples are models of vegetation~\cite{Prusinkiewicz1990}, cloudscapes~\cite{Webanck2018}, roads~\cite{Galin2010}, street networks~\cite{Parish2001}, and buildings~\cite{Mueller2006, Schwarz2015}.

\annot{scientifically-accurate procedural modeling}Procedural modeling has been utilized in sciences beyond visually-plausible modeling to create scientifically-accurate models. Biologists are recreating the mesoscale using procedural modeling methods, based on evidence from nanoscale and microscale measurements. Johnson et al.~\cite{Johnson2014} have proposed a system called cellPACK that takes as an input a so-called recipe, a description of how structures should be positioned in the organism model. A packing algorithm then iteratively places the macromolecular building blocks into different compartments of the organism. This compartment is described by a discretized distance volume, which is packed and updated in a sequential manner. Currently, on a desktop workstation, such a packing process takes several minutes up to hours, to pack a representation of the HIV particle 100~\si{\nano\metre} in diameter. The specification of the recipe, however, is a human-readable textual rule definition that relies on the accurate specification from the user. In our approach, instead, the rules are learned from the user interaction during interactive and intuitive modeling of few structural elements directly in the 3D environment. Such elementary 3D models of nanoscale elements can---in spirit of procedural modeling---be assembled into much larger mesoscale complexes that can be interactively constructed and visualized.

\section{Method}
\begin{figure*}[t]
	\centering
	\includegraphics[width=\linewidth]{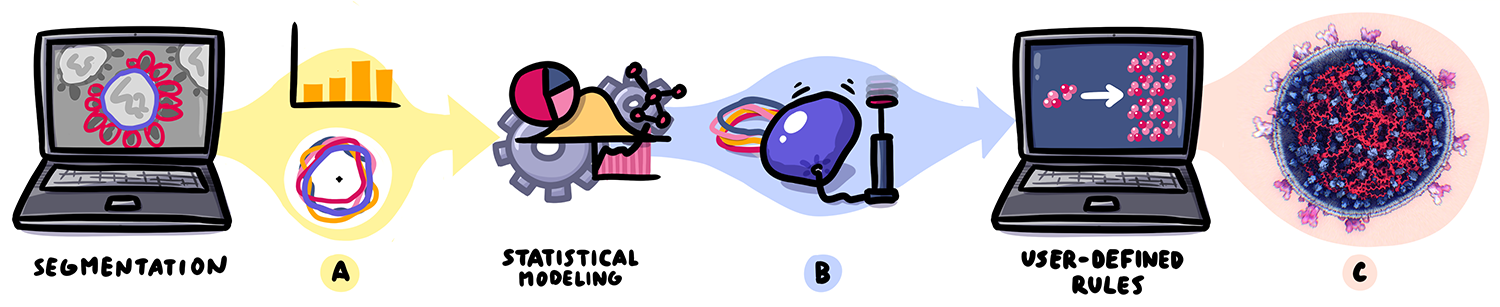}
	\caption{An overview of our mesoscale  modeling pipeline. First, user specifies the segmentation of the contours and visible membrane-embedded proteins. Contours and a histogram of amounts of proteins within individual parts of the membrane (a) are used for statistical contour modeling that is inflated into 3D mesh and  populated with membrane proteins (b). Subsequently, user specifies rules of how invisible proteins should be placed within the 3D model. The output is the finalized model (c), which can be iteratively refined by modifying the rules.}
	\label{fig:overview}
\end{figure*}
\annot{hint that this is not out of scope for scivis}In scientific data visualization, we use procedural modeling methods that result in plausible representations not only for a broad audience, but also for scientists for hypothesis generation, testing, or even in the simulation of stability and dynamics. To be able to create models that are scientifically relevant, our modeling framework needs to allow for versatile structural arrangement specification and needs to support integration with acquired evidence from microscopy data. This requirements differ from already introduced procedural modeling methods. For example L-system models~\cite{Lindenmayer1968} are typically topological trees and the procedural models are based on growing plants according to the tree structure or the architecture where models are generated by top-down subdivision with elements in regular arrangement. Mescoscale biological models are different case. We have many elements that are in relation to each other and interact with each other. Also arrangement is much more irregular than architecture. Therefore, we do not go the route of domain specific languages~\cite{Jiang2020}, but aim to design an interactive procedural modeling framework to tackle with biological models.

\annot{inputs}The mesoscale biological structure is typically characterized at the nanoscale by its molecular composition, where molecular structure can be either measured or simulated. The microscale is characterized from microscopy images or tomographic volume reconstructions of the entire entity. Rough shapes of the macromolecules can often be observed in these image data, so typically several hypotheses can be formulated about the specifics of the assembly. Usually the membrane boundaries and associated proteins are more recognizable than the soluble assemblies inside the membrane. Therefore we model the membrane information based on image data and the information inside the membrane is characterized through interactive 3D modeling using structural rules.

\annot{membrane data}In our work we concentrate on the extraction of membrane contours that are often obviously revealed in microscopic images, which are usually used to show scientific observations. First, a handful of membrane contours are traced by the user. These contours are co-registered to analyze their variation. Such representation is statistically captured so that many new contours, similar to the input samples, can be generated. Based on the contour information, a three-dimensional particle geometry is estimated that matches the contour shape. Resulting mesh representation of particles are populated with molecules bound to the membrane according to the observations in the images. We characterize the molecular distribution around the contour and estimate a corresponding distribution for the entire particle surface.

\annot{no-data scenario}Once the information from the images is incorporated into the mesoscale model, further modeling of elements that are not directly observed in image data is used to complete the model. Several hypotheses can be generated to express what a biologist considers as a valid assembly configuration. The modeling proceeds through an interactive process, where the modeler expresses certain spatial relationships on exemplary structural representatives and an interactive 3D visualization shows how this rule is applied for the corresponding molecular population. Based on the instantaneous visual feedback, the modeler can revise previous inputs to obtain the desired assembly. In this stage hierarchical relationships can be utilized for expressing the rules that define distance and orientation distributions among molecular instances. The scene population is corrected by collision handling so that a valid molecular scene results from the application of the rules.

The overview of the modeling process and involved steps described above are shown in \autoref{fig:overview}. The following sections (\autoref{sec:learning}, \autoref{sec:rule}) describe the technical details of our approach. 

\section{Learning Shapes and Distributions from Imaging}
\label{sec:learning}
\begin{figure}[b]
    \centering
    \includegraphics[width=0.6\linewidth]{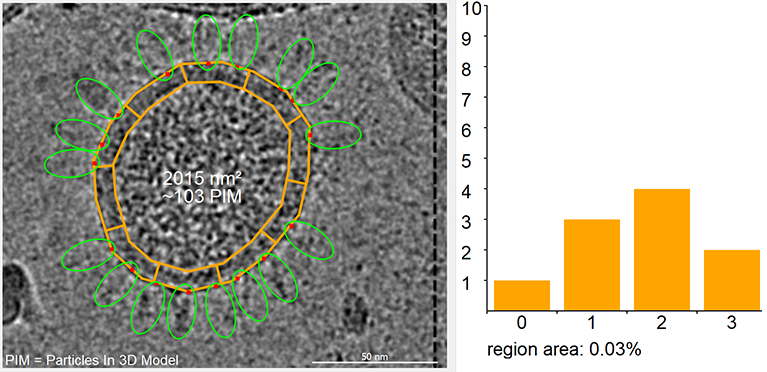}
    \caption{Input electron microscopy~\cite{Liu2020} image after segmentation. Left: The contour with a band is created. Right: The histogram representing the number of particles per contour band region.}
    \label{fig:dist-widget}
\end{figure}
\begin{figure*}
    \centering
    \includegraphics[width=0.17\textwidth]{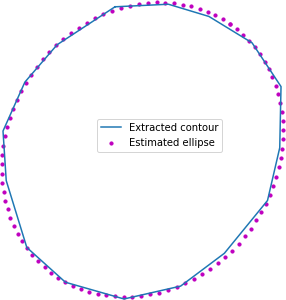}
    \includegraphics[width=0.44\textwidth]{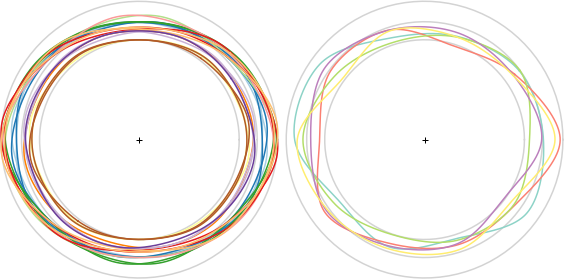}
    \includegraphics[width=0.21\textwidth]{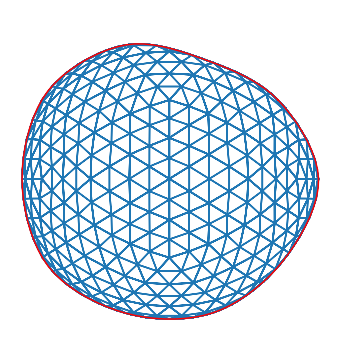}
  \includegraphics[width=0.40\textwidth]{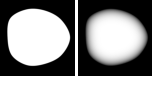}
    \includegraphics[width=0.42\textwidth]{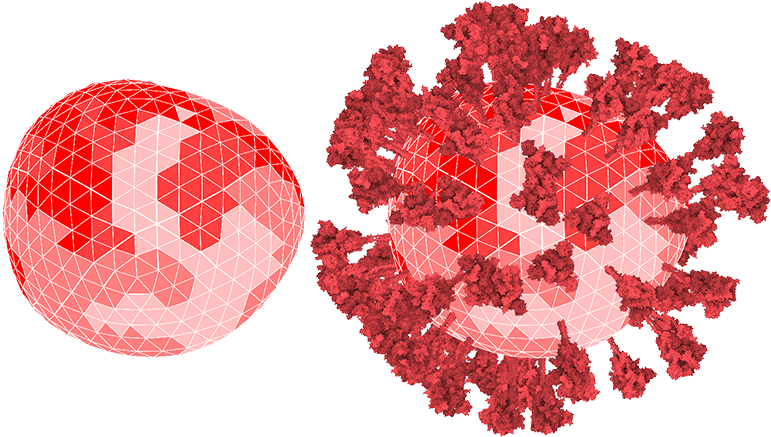}
    \caption{Statistical contour modeling for particle mesh generation: The contour is approximated by an ellipse that is used for bringing all contours into a canonical form (top left). Statistical contour model is generated from set of contours and new contours can be generated (top middle two). Newly generated contour is rasterized for contour inflation (bottom left two). Two dimensional mesh is generated (top right) which is then inflated into 3D object and populated with spike protein amounts per triangle (bottom right).}
    \label{fig:statistical}
\end{figure*}
\annot{rapid segmentation}For rapid processing of electron microscopy images, we implemented a segmentation tool that produces the input for learning the contour of the membrane and the protein distribution on the membrane. The user creates an outline, the outer contour of the cell, and places small elliptical proxy objects representing proteins scattered over the surface of the cell. The major axis of the ellipse is aligned with the main axis of the protein. We perform this quick feature extraction for all proteins that are close to the cross-section or silhouette of the membrane as shown in \autoref{fig:dist-widget}. These proteins naturally are not exactly on the contour, but they are located close to it within a certain surface band. To characterize this band, an inner contour is specified. The user can easily specify the thickness of this band on which the marked surface proteins are located.
With this we have fully characterized the most important aspects of the input microscopy image. Next a distribution of the surface proteins on the membrane band needs to be estimated. For this, we subdivide the band into equally sized surface patches and count the amount of proteins associated with each patch. To characterize the distribution of the proteins from what we see on the membrane contour, we store the per-patch protein counts in a histogram. This gives us a distribution  function of amounts of protein per patch area which we use when we populate membrane-protein instances on the 3D model of the membrane.

\annot{ellipse fitting}Once we obtain a set of membrane contours extracted from multiple virion particles, we then use them for generating new distinct contours that have similar characteristics. For that we need to register all contours into a common coordinate system. We do this by fitting an ellipse to each contour and then translating and rotating the contours such that the approximating ellipses are in canonical form. We assume that all contours can be approximated by an ellipse. To parametrize an ellipse, we use the two focal points and the semi-major length. A point $\vect{p}$ is on the ellipse if and only if: 
\begin{equation}\label{eqn:ellipse}
\left \| \vect{p} - \vect{c_{1}} \right \|_{2}+ \left \| \vect{p} - \vect{c_{2}} \right \|_{2} = 2a
\end{equation}
where $\vect{c_{1}} = (c_{1}.x, c_{1}.y), \vect{c_{2}} = (c_{2}.x, c_{2}.y)$ are the focal points and $a$ is the semi-major length. To fit an ellipse to a set of data points ${\vect{p_{i}} = (p_{i}.x, p_{i}.y)}_{i = 1}^{n}$, we pose the problem as optimization problem: 
\begin{displaymath}\label{eqn:optimize}
\min_{\vect{c_{1}} , \vect{c_{2}}, a} \frac{1}{n}\sum_{i=1}^{n}(\left \| \vect{p_{i}} - \vect{c_{1}} \right \|_{2}+ \left \| \vect{p_{i}} - \vect{c_{2}} \right \|_{2} - 2a)^{2}
\end{displaymath}

This objective function has a global minimum at infinity. When the two focal points move to infinity and the semi-major length tends to infinity, the value of this function approaches zero. We therefore add an L2 regularizer to avoid the undesirable global minimum at infinity: 
\begin{equation}\label{eqn:optimize_penalty}
\underset{\vect{c_{1}} , \vect{c_{2}}, a}\min \left [ \frac{1}{n}\sum_{i=1}^{n}(\left \| \vect{p_{i}} - \vect{c_{1}} \right \|_{2} \\+ \left \| \vect{p_{i}} - \vect{c_{2}} \right \|_{2} - 2a)^{2} + \frac{\lambda}{n}(\vect{c_{1}}^{2} + \vect{c_{2}}^{2} + (2a)^{2}) \right ]
\end{equation}
 where $\lambda$ is a tuning parameter.
In the initialization, $a$ is initialized as the mean of the distance from the data points to the mean of all data points $\vect{p_{\mu}}$, $\vect{c_{1}} = (-\frac{a_{max}}{2}, 0), \vect{c_{2}} = (\frac{a_{max}}{2}, 0)$, where $a_{max}$ is the largest distance from a data point to $\vect{p_{\mu}}$. After initialization, the penalized objective function \autoref{eqn:optimize_penalty} can be solved by gradient descent. The obtained ellipse has semi-major length $a$, semi-minor length $b$, center $\vect{c_{e}}$ and angle of rotation $\vect{\theta_{e}}$. The example of estimated ellipse can be seen in \autoref{fig:statistical} top left.

\annot{Registration}To register the contours into a common coordinate system, we estimate the translation and rotation based on the approximating ellipse. First, the segmented contour is translated to the origin $\vect{O}(0, 0)$ by translation vector $\vect{t} = -\vect{c_{e}}$. Then, the segmented contour is rotated by angle $-\vect{\theta_{e}}$. 
\annot{normal ninja}Each contour in the set of contours is reparameterized by $\vect{N_{p}}$ points $\vect{p_{i}}$. Each point
$\vect{p_{i}}$ is defined by an angle $\vect{\theta_{i}}$ and a distance $r_{i} = \left | \vect{O}\vect{p_{i}} \right |$ in a polar-coordinate system. To get more data from the contours, we create 3 augmented contours for each contour. The first augmented contour is obtained by a rotation with angle $\pi$. The second and third augmented contours are obtained by flipping the original and the rotated contour through the x-axis. To generate a new contour from the contours, we compute a per-angle one-dimensional normal distribution by casting a ray from the origin $\vect{O}$ in all $\vect{\theta_{s}}$ directions and intersecting all contours with this ray. To the $\vect{N_{p}}$ intersection ($\cap$) points $\vect{p_{\cap_{i}}}$, we fit a normal distribution for $\vect{N_{p}}$ $\vect{r_{\cap_{i}}} = \left | \vect{O}\vect{p_{\cap_{i}}} \right |$ with mean $\vect{\mu_{s}}$ and standard deviation $\vect{\sigma_{s}}$ as parameters. We also truncate the normal distribution to the minimum and maximum distance values in the data. From these parameters, we perform rejection sampling of the truncated normal distribution of $\vect{r_{\cap_{i}}}$ for each angle $\vect{\theta_{s}}$. Finally, we interpolate the points using Catmull-Rom splines to create a new contour. The input contours and a number of generated contours are shown in \autoref{fig:statistical} top middle two images. The algorithm is listed in \autoref{alg:generate-contours} and \autoref{alg:rejection-sampling}.

\annot{generating a virus particle}From the learned contour our next step is to generate a membrane, which is a three-dimensional ellipsoidal \emph{potato}-like object. Our object should feature the same shape characteristics as observed on the particle's silhouette or cross-section, depending on the imaging method, termed here as the contour. We model the object based on three principal dimensions, $d_{1} \geq d_{2} \geq d_{3}$ of an ellipsoid and characterize it by two aspect ratios, namely as elongation index $EI = d_{2}/d_{1}$ and flatness index $FI = d_{3}/d_{2}$~\cite{Torpa2006}. $EI$ can be determined from the contour, $FI$ is defined by the user then it is used for determined $d_{3}$. 
Next, we need to extrude the contour into three dimensions. For this task, we employ the standard contour inflation method from sketch-based modeling~\cite{Williams1991}.
To assign a depth ($z$) value to all points on the three-dimensional object we proceed as follows. First, we make a binary mask from the contour so that $0.0$ is assigned to the outside and $1.0$ to the inside of the shape. Then, we apply a cascade of Gaussian filters (with radius 32, 16, 8, 4, 2, 1) on the image mask. After each smoothing pass, the resulting image is multiplied with the original image mask, so that all pixels that are outside the contour are again set to zero. The resulting image is used for assigning the depth values symmetrically on both subspaces partitioned by the contour plane of $z=0$ (see \autoref{fig:statistical} bottom left two images for an example). 
The next step is to create the three-dimensional object represented by a triangular mesh. We create a 3D sphere with approximately equally-sized triangles where the radius is the largest radius from all contour points to the origin $\vect{O}$. After that, we project this mesh onto the $z=0$ contour plane. This projected mesh is distorted to the shape of the contour. The example 2D mesh backprojected onto the contour plane can be seen in \autoref{fig:statistical} top right. Finally, for each mesh point, we extrude its $z$-coordinate to \emph{inflate} the contour. The $z$-coordinate value of each point of the mesh is calculated as a multiplication of half of $d_3$ with the corresponding pixel value (with the same $x,y$ coordinates) from the previously calculated depth image in \autoref{fig:statistical} bottom left. 

\annot{distributing membrane proteins}In the image segmentation phase a band around the contour was created. This band was subdivided into equally sized segments. Moreover, each of these segments was evaluated by a number representing the amount of membrane proteins belonging to the segment. With this construction, we obtain the distribution of the number of membrane proteins per given area. We use this distribution for populating the membrane proteins on the triangular mesh. The membrane protein density of triangles is computed in the following way: The 3D mesh is partitioned into approximately same-sized triangular patches. The size of the patch is determined by the size of the area of segments of the bands from the 2D contour. Afterwards, every patch is associated with one value from the above distribution. We use random sampling of the distribution. Then we distribute the number of membrane proteins among the triangles that belong to the current patch. Examples of generated 3D meshes with associated protein counts per triangle can be seen in \autoref{fig:statistical} bottom right along with the examples of membranes with the membrane proteins.

\section{Interactive 3D Rule Specification}
\label{sec:rule}

\annot{no-image scenario intro, two building elements} The second part of our approach is used to populate model with biological elements that are placed in relation  to other elements in the cell. While some of these elements cannot be clearly seen in EM images, their structural information is generally understood or there is at least a hypothesis about the structural organization. For example a protein can be in a spatial relation (position, rotation) to another protein. The rules encode how new elements can be placed based on the geometry of already existing elements.


\annot{building blocks definition} We create a three-dimensional \emph{model} that consists of a set of \emph{elements}. Our interactive procedural modeling approach organizes the elements in a tree. An element consists of the following: 1) A name to identify the element, e.g. to select an input element to a rule. 2) A type that can be either auxilliary or instance. An auxialliary element will be invisible in the final model and an instance will be visible. We often refer to an auxilliary element as \emph{skeleton}. 3) The element geometry that can be either a polygonal mesh, a poly-line, or a set of points. Sometimes, the geometry is only a single polygon, line segment, or point. 4) An oriented bounding sphere that consists of a local coordinate system used to position the geometry in the world coordinate system and three scale factors to determine the size of the element.

\begin{figure}
    \centering
    \includegraphics[width=0.5\linewidth]{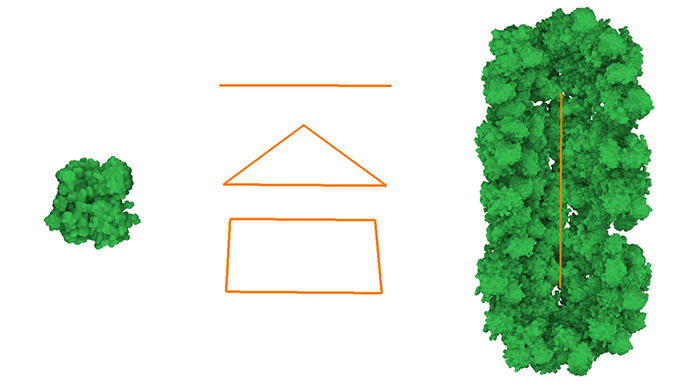}
    \caption{Illustration of element geometry. Left: protein instance from a database. Middle: a line segment, triangle, and rectangle. Right: an arrangement of protein instances around a line segment.}
    \label{fig:building-blocks}
\end{figure}

We use a library of structural models, e.g. proteins, in our framework. Many of these models are freely available on the internet. The most common form in which they are distributed is a list of atoms where the type and position of each atom is specified. Conceptually, we could convert these descriptions into 3D meshes, but we typically keep them in a different representation (e.g. set of spheres) for faster rendering. We also assign an identifier $G_{id}$ to them. These identifiers will be used in the rules to specify the geometry of elements.
We also use a library of elementary meshes, such as single polygons, a tetrahedron, or an icosahedron that proved to be useful as auxiliary geometry.
See~\autoref{fig:building-blocks} for an illustration of example geometries.

\subsection{Definition of rule}
The main function of a rule is to identify an element in the current model and to create one or multiple elements either as children or as siblings in the derivation tree. In contrast to other popular procedural modeling systems, e.g.~\cite{Mueller2006,Prusinkiewicz1990}, our rules are not described by a script like language, but they are designed and executed in an interactive editor. The user can interact with elements in a CAD-like environment, e.g. positioning and rotating elements using a virtual gizmo tool.
In the following, we will describe the most important concepts and parameters. We plan to release the executable and detailed UI documentation upon acceptance.



\subsection{Derivation of the model}
The derivation of a model starts with an auxiliary root node of the tree and an empty model. Elements are placed by processing the specified rules and rule groups in a sequence. The user has full control of the rule execution and can execute all rules at once, execute rules step by step, and perform interactive edits between the execution of rules. Further, the user can undo rules or even partially undo rules.
Rules take elements currently presented in the scene ( identified by their name) as input and generate zero, one, or multiple new elements.

If rules $r_i$ are placed in a group they can be either applied in an alternating manner, or rules can be selected randomly among the set of rules in the rule group according to their probability $r_i.probability$.

\annot{probability functions} Several geometric parameters can be specified as constants or as probability distributions. A probability distribution can be modeled by combining Gaussian and uniform functions as building blocks. The values are automatically normalized so that their sum integrates to one.

\annot{element rotations} In all the rules, several rotational variants can be used to specify a transformation. Currently implemented variants are: user-defined rotation, random rotation, normal vector orthogonal to a parent element normal vector and element normal vector aligned to a parent normal vector. Moreover, these rotations can be extended by user-specified yaw, pitch, roll distributions.

\begin{figure*}[t]
	\begin{minipage}{.48\textwidth}
		\centering
		\includegraphics[width=.9\linewidth]{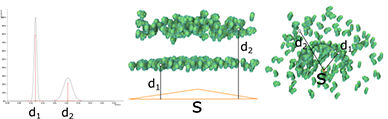}
		\caption{Distance rule illustration. Left: The definition of a probabilistic distance function. Middle: Application of the rule to a triangular skeleton. Right: Application of the rule to a point skeleton.}
		\label{fig:rule-master-slave-distance}
	\end{minipage}
	\hfill%
	\begin{minipage}{.48\textwidth}
		\centering
		\includegraphics[width=.9\linewidth]{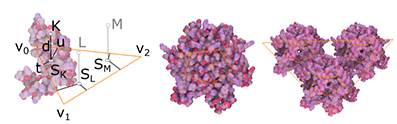}
		\caption{Parent-child Rule. Left: Illustration of a relative rule created on a triangular skeleton. Right-Top: The same rule applied to a pentagonal skeleton. Right-Bottom: The pentagonal skeleton model is bound to a triangular skeleton.}
		\label{fig:rule-relative}
	\end{minipage}
	\hfill
	\begin{minipage}{.48\textwidth}
		\centering
		\includegraphics[width=.9\linewidth]{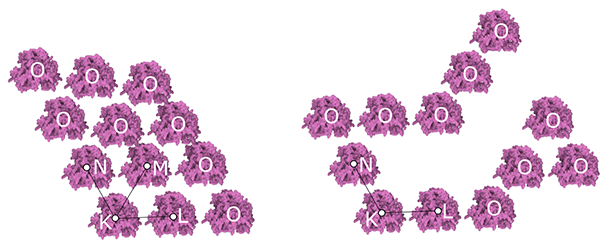}
		\caption{Siblings rule. Left: all three transformations are applied in every iteration. Right: One transformation is selected randomly per iteration.}
		\label{fig:rule-siblings}
	\end{minipage}
	\hfill%
	\begin{minipage}{.48\textwidth}
		\centering
		\includegraphics[width=.9\linewidth]{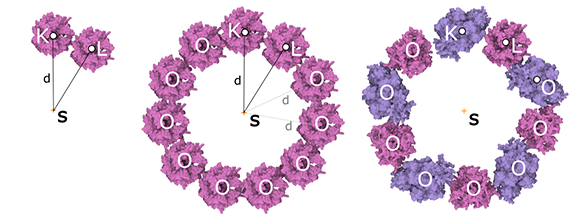}
		\caption{Siblings-parent rule. Left: Creation of a rule. Middle: Application of the rule. Right: Two rules applied alternatively to place two different elements in a circle.}
		\label{fig:rule-siblings-parent}
	\end{minipage}
	\hfill
\end{figure*}

\annot{collision detection} An important part of our approach is collision detection. We implemented an OctTree accelerated method. A naive collision detection algorithm turned out to be unusable due to the amount of elements in the model (~200000 element instances). 
We use an OctTree with 4 levels of subdivision. Every element in the scene is assigned a bounding sphere. This bounding sphere can be additionally scaled by the user to approximate the object better when the object is long but thin. Although this can lead to overlapping of elements, the property can be exploited for example in cases where string-like elements are placed in a plane close to each other. The user can control how close elements can be before creating a collision.
Once a new element candidate is generated all leaf octants of the tree intersecting the element's bounding sphere are fetched and used for collision detection. If there is no collision, the element is created. Otherwise, the element is not created. To avoid rules that are not terminating because of collisions, we employ a parameter $collisions_{max}$ to specify the maximum number of consecutive detected collisions. If that number is reached, a rule is terminated.

\annot{rule reuse} The specifications of all rules are stored in a file. Thus, the user can create a library of rules that can be then re-used as templates to build other mesoscale models.

\subsection{Type of rules}
We identify and implement four main classes of rules: parent-child, siblings, siblings-parent, and connection rule.

\subsubsection*{Parent-child rule}
\label{sec:master-slave}
In this rule new child elements are added to a parent element with name $Name_{in}$ given as input.
We employ two types of rules called distance rule and relative rule.

\annot{distance rule} The main purpose of the distance rule is to create new elements in a specified distance to the parent. The distance $d$ can be either a constant or modeled probability distribution that is sampled each time a new element is created. To determine the position of the new element, a random point on the parent geometry is generated and translated along the normal vector according to the (sampled) distance $d$. Another parameter determines if the translation happens along the positive normal direction, negative normal direction, or randomly selected among the two.
In~\autoref{fig:rule-master-slave-distance} left we illustrate a probability functions that is modeled as a combination of two Gaussians with mean $d_1$ and $d_2$. The Gaussian around $d_1$ has a higher weight than the Gaussian around $d_2$. The resulting population of elements using a triangle and a point skeleton as parent are presented in~\autoref{fig:rule-master-slave-distance} middle and right, respectively.

\annot{relative rule} The relative rule specifies the location of new elements with respect to a vertex of a polygon of the input element.
For example, in \autoref{fig:rule-relative} left)
a position $K$ is specified by the user and subsequently encoded with respect to vertex $v_0$. The position is computed by the parameters $t,u$ that specify the distance from the edges connected to $v_0$ and the distance $d$ along the normal of the polygon. The parameters $t,u$ specify the location of $S_K$, the closest point on the polygon.
From these rule parameters, the corresponding positions positions $S_L$ and $S_M$ can be found inside the triangle and new elements are placed in the points $L$ and $M$ that are in the distance $d$ along the normal vector from positions $S_L$ and $S_M$, respectively.
The rule created from the previous process can be transferred and applied to any polygon, e.g. to create a pentamer as shown in~\autoref{fig:rule-relative} middle. 
The example in~\autoref{fig:rule-relative} right shows two subsequent applications of the rule to model the stucture of a viral capsid. First, the relative rule is used to create three pentagon elements as children of a triangle element, Second, proteins are created as children of each of the pentagons.

\subsubsection*{Siblings rule}
\label{sec:siblings}
The siblings rule creates new elements and adds them as siblings to the same parent in the tree.
The most important parameter of the siblings rule are a set of transformations $T_i$. These transformations are typically a combination of translation and rotation. Each transformation also has an associated probability $T_i.prob$.
To apply the rule, a transformation $T_i$ is selected according to the probabilities $T_i.prob$.
Then, a new element with name $Name_{out}$ is generated by transforming the coordinate system of the input element $Name_{in}$ and setting the geometry as specified by the identifier $G_{id}$.
A parameter $T_{num}$ determines how many transformations will be selected. If $T_{num}$ is equal to the number of transformations, all transformations will be selected and the probabilities will be ignored.
The rule is invoked recursively for newly created elements. Identical to previous rules, the parameter $count_{max}$ determines how many elements are inserted.
In~\autoref{fig:rule-siblings} left, three different transformations $T_1 = K \rightarrow L$, $T_2 = K \rightarrow M$ and $T_3 = K \rightarrow N$ were created. The user can generate these transformations interactively. In this example, an additional 8 instances labeled $O$ were generated recursively (for $T_{num} = 3$). 
In the example~\autoref{fig:rule-siblings} right, $T_{num}$ is set to one to showcase the random selection of transformations.

\subsubsection*{Siblings-parent rule}
\label{sec:siblings-parent}
The siblings-parent rule is an extension of the siblings rule. The user specifies a transformation to the sibling element with name $Name_{in}$ as before. After applying the transformation, the new element $Name_{out}$ is snapped to a given distance $d$ to the parent shape of $Name_{in}$. This distance preservation acts as correction factor. 
The main benefit of this rule is that the user who wants to distributes elements in a circle around a point or a spiral around a line segment does not have to precisely measure the angle and translation.

In~\autoref{fig:rule-siblings-parent} left, the parent shape is a point labeled $S$ and the input element is labeled $K$ and the newly generated element is labeled $L$. The transformation $T = K \rightarrow L$ consists of a translation. In subsequent applications of the rule the distance $d$ to the skeleton $S$ is preserved (see~\autoref{fig:rule-siblings-parent}  middle). 
The user can specify a group of rules that can be applied in iterations. In \autoref{fig:rule-siblings-parent} right two rules with transformation $K \rightarrow L$ and $L \rightarrow M$ were created and applied.

\subsubsection*{Connection rule}
\label{sec:connection}
This rule is used for creating string-like structures that connect a given set of 3D points. This set of points is typically generated by other rules. The output of this rule is a polyline element. The rule proceeds in three steps. First, an initial polyline is created by connecting the 3D points. For this purpose, the generator starts at a random point and connects it to a random one in its close proximity until all points are connected. The resulting polyline may have strong kinks. To remove the kinks, a cubic interpolation and subdivision is applied resulting in a smoother polyline. Fiber structures, that we would like to model, are characterized through a persistence length property that expresses the bending stiffness of a fiber. Midpoint displacement is able to incorporate the target stiffness by increasing or decreasing the amount of displacement~\cite{Klein2019}. Therefore, in the third step, the midpoint displacement algorithm is used to enhance the curve with detailed windings.

\begin{figure}[t]
    \centering
    \includegraphics[width=0.6\linewidth]{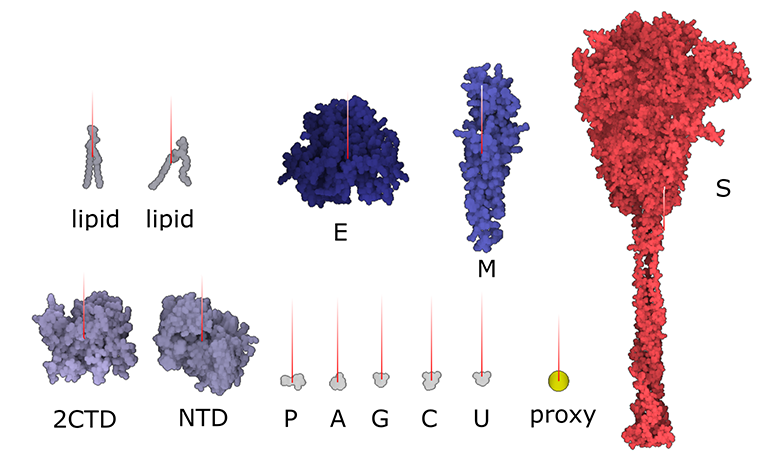}
    \caption{Setting of normal vectors to individual PDBs.}
    \label{fig:normals}
\end{figure}

\section{Use case: SARS-CoV-2}
The novel coronavirus SARS-CoV-2 is currently posing an international threat to human health. As with previous SARS and MERS outbreaks, it emerged through zoonotic transfer from animal populations. These types of emerging viruses pose a continuing threat, and the biomedical community is currently launching a widespread research effort to understand and fight these viruses. Understanding of the mesoscale structure will play an essential role in understanding the modes of interaction of these viruses with their cellular receptors and designing effective vaccines. We introduce the topic in biology language first and then describe our modeling strategy.

Our mesoscale models integrate a growing body of cryoelectron micrographic data on entire virions with atomic structures of the biomolecular components. SARS-CoV-2 contains four structural proteins, a single strand of genomic RNA, and a lipid-bilayer envelope \cite{Neuman2016}. Other non-structural and/or host proteins may also be incorporated into the virion—this is a topic of current study in the field and it is not addressed in these models. Three of the structural proteins are embedded in the membrane. The spike (S) protein extends from the surface and forms the characteristic spikes that give the viruses their crown-like shape as seen by electron microscopy. The spikes recognize cellular receptors and mediate entry of the virus into cells. The membrane (M) protein has an intravirion domain that interacts with the nucleocapsid protein and is involved in packaging the viral genome as the virus buds from the infected cell’s surface. The envelope (E) protein is a small pentameric complex that forms an ion pore through the membrane, which is thought to be involved in the process of budding, with only a small number of copies being incorporated into the virus. The viral genome is a single strand of RNA about 30,000 nucleotides in length, one of the largest genomes of RNA viruses. It is packaged by the nucleocapsid (N) protein, which coats and condenses the RNA strand. 

\begin{figure}[t]
	\centering
	\includegraphics[width=0.6\linewidth]{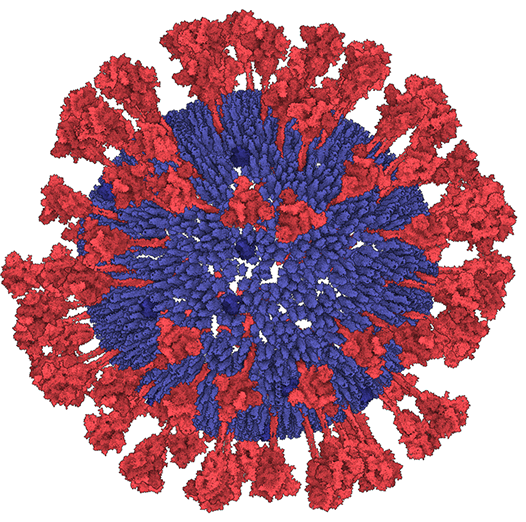}
	\caption{Spikes scattered on the surface of the 3D mesh according to the assigned amounts per triangle. Membrane and envelope proteins are uniformly distributed on the membrane.}
	\label{fig:mesh_m}
\end{figure}
At initiation of this project, structural information about SARS-CoV-2 proteins was limited, so we employed homology modeling to generate atomic structures of the viral proteins. The automated pipeline described here is created so that the most current structural models may be substituted as additional information is obtained. The project also made extensive use of the rapid modeling capabilities, to incorporate new sources of data as they became available. 

Several structures of the S protein in its prefusion state have been determined with cryoEM (PDBids: 6VSB, 6VXX, 6VYB). These structures cover most of the protein, leaving out about 130 residues at the C-terminus, corresponding to HR2, transmembrane and cytoplasmic domains. The N protein has been partially determined with X-ray crystallography (PDBids: 6VYO, 6YI3, 6WJI, 6M3M). As for SARS-CoV-2 M and E protein, none of them have been experimentally resolved yet. In this work, we took advantage of SARS-CoV-2 sequence homology with other betacoronaviruses to perform homology modeling and to obtain structural model for each protein component present in the SARS-CoV-2 virions. 

The S protein model was obtained by a composite homology modelling approach. Residues 27-1146 were modelled with SWISS-MODEL \cite{Waterhouse2018} using 6VXX as template, in order to reconstruct all the missing loops that were not solved in the experimental structure. Residues 1147-1212, 1213-1242, 1243-1273 corresponding to HR2, transmembrane and cytoplasmic domain respectively were submitted separately to the homology modeling server for oligomeric structures GalaxyHomomer \cite{Baek2017}. Template-based modeling generated the HR2 structural model, utilizing the structure of MERS-CoV fusion core (PDBid 4MOD, 39\% sequence identity) as template, while transmembrane and cytoplasmic models originated from ab-initio modeling methods. Structures of the four components were aligned in PyMol \cite{PyMOL2015} and merged to obtain a unique model in Coot \cite{Emsley2010}.

Pentameric E protein was modelled with the protein structure prediction server Robetta \cite{Kim2004} in ‘comparative modeling’ mode. The NMR structure of E protein from SARS-CoV (PDBid 5X29) was used as template, as the two proteins share 94\% of their amino acid sequence. The structural model for M protein (residues 11-203) was predicted by DeepMind using AlphaFold prediction methods and freely distributed in response to the COVID-19 crisis \cite{Jumper2020}. Residues 3-10 were built onto the original model with Coot \cite{Emsley2010} and a N-glycosylation was added to residue N5 through GLYCAM-Web \cite{Singh2019}. 
\begin{figure}[t]
	\centering
	\includegraphics[width=0.6\linewidth]{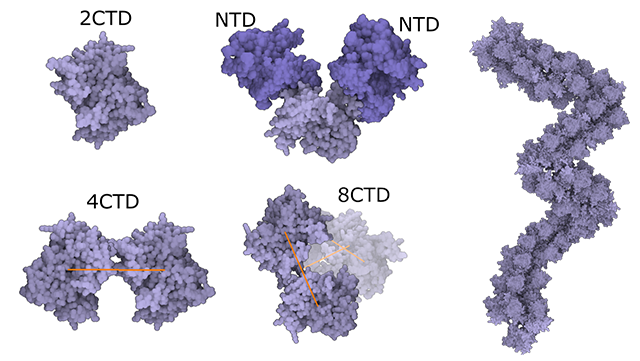}
	\caption{Rope-like N protein complex rule. Left: Creating an N protein CTD octamer structure and its relation with NTD. Right: Connecting N protein octamer structure to form the rope.}
	\label{fig:rope-ctd}
\end{figure}

The structure of the viral nucleoprotein complex is a topic of current speculation and study, so we created a model that is consistent with estimates of the number of copies of N, the length of the genome, and the locally-ordered arrangement of proteins as seem by cryoelectron microscopy \cite{Neuman2006}. N protein folds to form two stable globular domains, connected by a flexible linker. The C-terminal domain (CTD) forms dimers, as seen in PDBid 6WJI of the SARS-CoV-2 CTD. In the SARS CTD, these dimers may further associate into octamers as seen in PDBid 2CJR. The structure of the N-terminal domain (NTD) has been determined in PDBid 6YI3, and is often termed the RNA-binding domain, although RNA binds to multiple sites in the N protein \cite{Chang2014}. In cryoelectron micrographs, the nucleoprotein complex is seen to have short-range order as a close-packed collection of spherical features closely opposed to M protein on the inner face of the membrane \cite{Neuman2006}, and gently-isolated nucleoprotein complexes have the look of rope-like collection of beads roughly 15 nm wide \cite{Mcneughton1978}. These data have been integrated into a conceptual model where CTD octamers stack loosely into long ropes and NTD are loosely associated around the central rope, interacting with RNA \cite{Gui2017}. We have implemented this conceptual model, first building CTD octamers, then placing NTD, and finally building a continuous RNA strand that winds through the NTD. Explicit models for the intrinsically-disordered linkers and flexible portions of the N chain at the N- and C-termini will be a subject for future study.
In the rest of this section the modeling process is described. Firstly the model of virion is created. In the following section the process of RNA construction is described.

\subsection{Virion modeling}
\annot{prepation}Due to an arbitrary rotation of molecular models in PDB files, the user must assign a normal vector for membrane-bound components, to define the orientation and location within the membrane (see \autoref{fig:normals}). If this assignment is skipped, the default rotation, as stored in the data file, is used. The whole modeling phase is done taking into account the estimated amount of individual elements published in \cite{BarOn2020}.
\annot{2D segmentation}We implemented a tool in which the user can segment 2D electron microscopy images (see \autoref{fig:dist-widget}). Firstly, the scale of the image has to be set. We implemented a widget for selecting a portion of the image and assigning a real-world length. The tool then computes this scale and uses throughout the entire segmentation process. 
Afterwards, the user manually segments the image and creates the outer contour by drawing a polyline enclosing the virion. After the outer polyline is done, the inner contour is created by scaling down the outer contour. The scaling is driven by the user and can be updated whenever necessary. Once the inner contour is defined, the band between outer and inner contour is automatically subdivided into 10 equally-sized regions. 
As the last step, the user visually identifies the spikes in the image and marks them using proxy objects available in our utility. The tool automatically identifies the closest virion for the particle and assigns the particle into the corresponding contour region. After this assignment, the histogram is updated. 
\annot{3D mesh}Outer contour and histogram are the input for the statistical learning (\autoref{sec:learning}). We process the data and estimate a new contour as described previously. From the newly generated contour we create a 3D triangular mesh and assign each of its triangles the amount of containing S-protein instances based on the histogram sampling (see \autoref{fig:overview}).
\annot{S distribution}In the next step S proteins are placed on the surface of the mesh. The user specifies the distance of the center point of a spike to the surface of the 3D mesh and the number of spikes to be placed. The parent-child rule \autoref{sec:master-slave} is used: the parent is the 3D mesh and children are the spikes. The illustration is depicted in \autoref{fig:mesh_m}.
\annot{M, E} A similar rule is used for both M and E proteins. The only difference is that these protein instances are uniformly distributed, \ie no amounts distributed over the 3D mesh based on observation on the contour is taken into account. These protein are not discernible on the images and their distribution is to date not characterized. For the time being our model assumes a uniform distribution on the mesh. For illustration see \autoref{fig:mesh_m}.

\annot{N proteins}The nucleoprotein complex is built in several steps. First, a fiber-like assembly of N protein conformations is built. The N protein conformation is modeled using CTD dimer (2CDT) and NTD instances as follows. Firstly, siblings rule \autoref{sec:siblings} is created to bind two NTD to 2CDT (see \autoref{fig:rope-ctd} top-left). Although there are only two relations depicted in the image, we create the total amount of 6 relations between 2CTD and NTD. In the population phase only two out of 6 created relations are randomly chosen. In the following step (see \autoref{fig:rope-ctd} bottom-left) 2CTD is bound to rotated 2CTD using siblings-parent rule \autoref{sec:siblings-parent} to a linear skeleton. This forms a tetramer 4CTD. Repeatedly, 4CTD is bound using linear skeleton to another instance of 4CTD forming an octamer 8CTD. The final N protein assembly is constructed using siblings-parent rule of 8CTD and a polyline skeleton. To create the polyline skeleton, we uniformly fill the interior of the 3D mesh by instances of a proxy object. This proxy object is a sphere that is customized to have a radius approximately the same size as the radius of bounding sphere of 8CTD. Applying connection rule \autoref{sec:connection} on such proxy spheres we obtain the polyline skeleton. Finally, siblings-parent rule is applied to 8CTD and the polyline forming the N protein assembly is created. RNA is then added to form the entire nucleoprotein complex, as described in the next section.

\begin{figure}[t]
	\centering
	\includegraphics[width=0.6\linewidth]{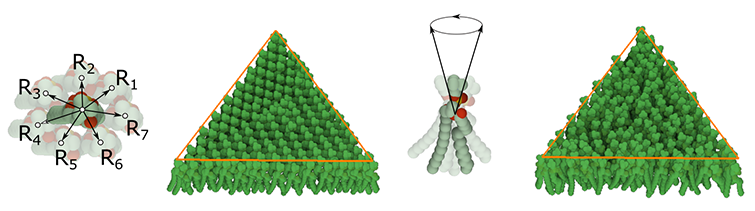}
	\caption{Population of lipids. Left: The rule with 7 relations is created for a lipid. Application of the rule on triangular skeleton forming patterns. Right: Modifying the rotation of the lipid by setting yaw, pitch, roll. Resulting population on the triangular skeleton.}
	\label{fig:model-lipids}
\end{figure}

\annot{lipid membrane}The lipid bilayer membrane is constructed using the siblings-parent rule from \autoref{sec:siblings-parent}. In reality, both layers can be modeled by the similar rule with the only exception that the lipids in one layer are rotated so that they are oriented to each other with their hydrophobic part. The construction of a single layer is as follows: Two copies of the same lipid models are added into the scene. The user creates a rule with several relations by translating (and rotating) one copy of the model to desired positions and stores these positions together with the distance to the 3D mesh parental skeleton. In our model, 7 relations $R_i$ are created (see \autoref{fig:model-lipids} left) for a transition from a lipid to another lipid. We use two models of different lipids - $Li_1$ and $Li_2$. Therefore, there are 7 relations for each of the combination: $Li_1 \rightarrow {Li_2}$, $Li_2 \rightarrow {Li_1}$, $Li_1 \rightarrow {Li_1}$. The generating process starts in randomly chosen triangle of the 3D mesh parental skeleton. In the beginning, one lipid is added. In the next steps up to 7 new lipids can be generated. The rule is re-applied on these new instances. This process continues until 100 consecutive collision hits are accumulated which indicates a fully dense lipid membrane. Populating models over the surface only by limited amount of relations would lead to an occurrence of visible patterns. This is tackled by fine-tuning the rotation of newly placed element by specifying how can standard deviation in yaw, pitch, roll vary (see \autoref{fig:model-lipids} right).

\subsection{RNA modeling}
\begin{figure}[t]
	\centering
	\includegraphics[width=0.7\linewidth]{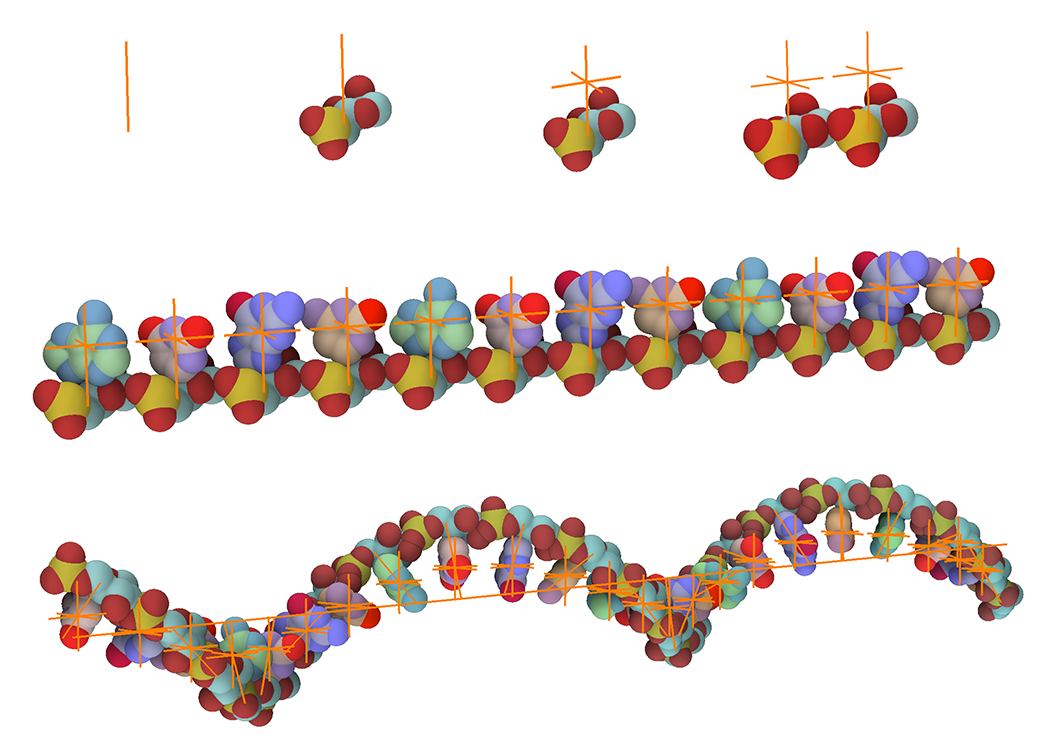}
	\caption{Illustration of RNA building. Top: Creation of a nucleotide and binding of two RNA nucleotides is illustrated (only the phosphate sugar backbone is shown, the bases are going to be populated at the point proxy above the backbone). Middle: The replication of the rule and replacing of proxies with A,C,G,U models. Bottom: The rule with a rotation incorporated applied to a line segment skeleton.}
	\label{fig:rna}
\end{figure}

We have modeled RNA using five elementary models: four RNA bases adenine (A), cytosine (C), guanine (G), uracil (U) and a model consisting of phosphate and sugar (P) that forms the RNA backbone. A model of an individual nucleotide is created by following approach. To a line skeleton using parent-child \autoref{sec:master-slave} rule a P part and a point skeleton are bound (see~\autoref{fig:rna} top). The point skeleton in this case has a role of a proxy object. In the population phase a corresponding model of a base (A,C,G,U) from a genome string is placed to this position to finish the formation of the intact nucleotide. The genome string is specified by the user during the definition of the point skeleton rule.

In the next step, the siblings-parent \autoref{sec:siblings-parent} rule of these individual nucleotides with a line skeleton is created. For simplicity, only translation is presented in~\autoref{fig:rna} top-right. Once this relation is applied to a line skeleton an RNA strand is created (see~\autoref{fig:rna} middle). 

If a rotation between two nucleotides is specified during the process of modeling, the the resulting RNA structure will twist along the line skeleton (see~\autoref{fig:rna} bottom). The consecutive bases with their phosphate sugar backbone forms the RNA string.  Now the RNA model is created as a template and can be used in any place where RNA is needed, it just needs a skeletal structure along which it will form the RNA backbone fiber and then populate the bases according to the given RNA sequence. In our case we replace the line skeleton with a polyline skeleton that represents a 3D curve connecting binding pockets of N proteins inside the core of the virion.

\begin{figure}[b]
	\centering
	\includegraphics[width=1.0\linewidth]{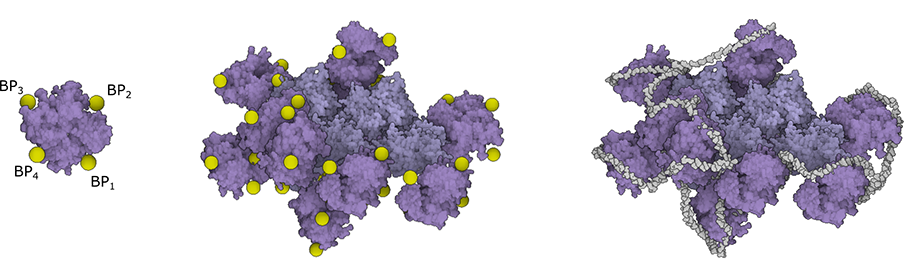}
	\caption{RNA proxy objects. Left: Specifying of proxy objects representing RNA binding pockets on the surface of a few N proteins. Middle: The binding pockets computed on all N proteins in the model. Right: The resulting RNA after populating A,C,G,U,P along the 3D curve approximating the proxy objects.}
	\label{fig:rna-n-proxies}
\end{figure}

RNA of the virion leads through predicted binding sites on the surface of N proteins. To specify the points through which the RNA should lead, a rule with relations of a proxy object $BP_i$ to N protein was created (see~\autoref{fig:rna-n-proxies} left). After populating the N proteins in the model, the algorithm computes positions of all proxies in the scene. 
The RNA backbone is obtained by using the connection rule  (see \autoref{sec:connection}) and a 3D curve generator.

\section{Discussion}
The virion model has been undergoing many revisions as the new information about its ultrastructure was updated in the literature. Using standard modeling approaches this would often lead to complete reassembly of the model. In our case few of the rules had to be redefined and an updated model was instantly generated. This real world experience of streaming new information has confirmed that our modeling framework is versatile  enough to accommodate for new revisions with a given set of rules, and that the modeling is a rapid process when it comes to complex structural characteristics of a virus particle. The benefit of the rule-based modeling is the nature of \emph{templating}  that becomes advantageous in a highly similar models of biological assemblies. Therefore, for example, once the RNA rules are specified, these can be effortlessly brought to another model. In case more structural knowledge comes in or a more advanced model of the RNA is refined, such model can be used in all mesoscale models which contain the RNA rule. The templating can be utilized for example in capsids, fibers, or membranes. This property inherently supports collaborative efforts, where one modeller revises initial models of the others and gradually a community can build a large base of mesoscale biological assemblies.

Today, availability of mesoscale models provides new opportunities for the understanding of SARS-CoV-2 structure and function. The number and distribution of spike proteins is still a matter of some conjecture, and is relevant to understanding of interaction of the virus with its cellular receptors and its interaction and neutralization by antibodies. The details of nucleoprotein condensation and packaging through interaction with the viral membrane protein are also of interest, since they provide possible targets for therapeutic intervention.

Currently, the whole approach is implemented as a single-threaded application on the top of the Marion molecular visualization framework~\cite{Mindek2018} and as a proof-of-concept implementation it is not significantly optimized for performance. Some processing stages are not calculated at an instance, however, we believe that the overall user experience is performant enough for rapid prototyping of biological mesoscale models. The resulting model consists of 100 S, 2000 M, 25 E, ~1000 N (in N-CTD and N-NTD), 29903 bp ss-RNA bases (GenBank: MN908947.3 \cite{Wu2020}) and 29903 P elements forming the RNA backbone, and ~180000 lipids. The entire model is created by 21 rules (with 57 relations) defined by the user. Several of these are different possible configurations of the same elements (as in the case of lipids). The population of S, E, M, N and all parts of RNA are processed within 2 seconds each. The population of lipids is the most computationally demanding part of the algorithm, it takes approximately two minutes for each inner and outer membrane. The main reason is that there are many lipid samples that are regressed. However, this time is heavily dependent on the rules defined. We have created very dense distribution of lipids with 7 relations for the rule. 

\section{Conclusion and Future Work}
In this paper, we have presented a new system for rapid modeling of mesoscale biological models. Challenged with frequent revisions of the SARS-CoV-2 model, the framework demonstrated its versatility and was able to incorporate any new structural insight. The result for science is two fold: a new technology and a new structural model of SARS-CoV-2\footnote{available at \url{nanovis.kaust.edu.sa/sars-cov-2-virus-model/}} that might lead to ideas for effective vaccination or treatment strategies. There are several research questions that are difficult to answer without an explicit imaged evidence. One such question could be: How many copies of RNA can a single virion pack? Is it just one or is there possibly enough space to accommodate another copy? The utility potential for hypothesis generation of biological questions that are of integrative structural nature is tremendous.

The framework allows for varying levels of accurate model specification. A model can be specified exactly so that one amino acid interacts with another one, or it can be placed more roughly. The system of rules preserves the accuracy which is given as input. Combined with the specification of flexibility and collision handling we can achieve even simple geometric docking. 

Our modeling of the microscopic scale of the membrane, is currently limited to simple star-shaped structures, \ie, there is a point inside from which all contour points are directly \emph{visible}. This is sufficient for simple virion shapes, however more complex shape modeling strategies would need to be employed for more complex shapes, such as the inner mitochondrial membrane, or Golgi apparatus for example. Another limitation is the case of \emph{sticky} fibers such as single-stranded genome macromolecules. These often form complex secondary structures that are enabled through sequence-interval complementarity. It is unclear whether such characterization could be expressed through our system, or we would need to expand the rule set.

It could be interesting to study the interplay of rule-modeled structures and reconstructing details from microscopic images that are hardly discernible by fitting a particular rule-expressed pattern into the image. Parallelizing our implementation would allow for interactive performance, where a large amount of conformations could be tested within a short time. Iteratively fitting the detail to an unclear microscopy image might be a way to solve an inverse problem in a brute-force manner.


Modeling with mouse interactions is not the only way, and with advanced speech recognition, a voice-controlled modeling becomes a possibility. Simultaneously, the ontology community has created a rich categorization of shapes, that have however no associated geometry. By establishing such association, through usage of terminology in shape ontologies, a verbal specification of models can lead to a desired model.


\bibliographystyle{unsrt}

\bibliography{template}

\begin{thebibliography}{10}

\bibitem{Baek2017}
M.~Baek, T.~Park, L.~Heo, C.~Park, and C.~Seok.
\newblock Galaxyhomomer: a web server for protein homo-oligomer structure
  prediction from a monomer sequence or structure.
\newblock {\em Nucleic Acids Research}, 45(W1):W320--W324, 04 2017. doi: {{%
10\hspace{.1pt}\discretionary{.}{%
}{.}\hspace{.4pt}1093\discretionary{/}{%
}{/}nar\discretionary{/}{%
}{/}gkx246}}


\bibitem{BarOn2020}
Y.~Bar-On, A.~Flamholz, R.~Phillips, and R.~Milo.
\newblock Sars-cov-2 (covid-19) by the numbers.
\newblock {\em eLife}, 9, 03 2020. doi: {{%
10\hspace{.1pt}\discretionary{.}{%
}{.}\hspace{.4pt}7554\discretionary{/}{%
}{/}eLife\hspace{.1pt}\discretionary{.}{%
}{.}\hspace{.4pt}57309}}


\bibitem{Bhatia2019}
H.~Bhatia, H.~I. Ing\'{o}lfsson, T.~S. Carpenter, F.~C. Lightstone, and P.-T.
  Bremer.
\newblock {MemSurfer}: A tool for robust computation and characterization of
  curved membranes.
\newblock {\em Journal of Chemical Theory and Computation}, 15(11):6411--6421,
  2019. doi: {{%
10\hspace{.1pt}\discretionary{.}{%
}{.}\hspace{.4pt}1021\discretionary{/}{%
}{/}acs\hspace{.1pt}\discretionary{.}{%
}{.}\hspace{.4pt}jctc\hspace{.1pt}\discretionary{.}{%
}{.}\hspace{.4pt}9b00453}}


\bibitem{Blinn1982}
J.~F. Blinn.
\newblock A generalization of algebraic surface drawing.
\newblock {\em ACM Transactions on Graphics}, 1(3):235--–256, 1982. doi: {{%
10\hspace{.1pt}\discretionary{.}{%
}{.}\hspace{.4pt}1145\discretionary{/}{%
}{/}357306\hspace{.1pt}\discretionary{.}{%
}{.}\hspace{.4pt}357310}}


\bibitem{Bruckner2019}
S.~Bruckner.
\newblock Dynamic visibility‐driven molecular surfaces.
\newblock {\em EG Computer Graphics Forum}, 38(2), 2019. doi: {{%
10\hspace{.1pt}\discretionary{.}{%
}{.}\hspace{.4pt}1111\discretionary{/}{%
}{/}cgf\hspace{.1pt}\discretionary{.}{%
}{.}\hspace{.4pt}13640}}


\bibitem{Chavent2011}
M.~Chavent, A.~Vanel, A.~Tek, B.~Levy, S.~Robert, B.~Raffin, and M.~Baaden.
\newblock Gpu-accelerated atom and dynamic bond visualization using hyperballs:
  A unified algorithm for balls, sticks, and hyperboloids.
\newblock {\em Journal of Computational Chemistry}, 32(13):2924--2935, 2011.
  doi: {{%
10\hspace{.1pt}\discretionary{.}{%
}{.}\hspace{.4pt}1002\discretionary{/}{%
}{/}jcc\hspace{.1pt}\discretionary{.}{%
}{.}\hspace{.4pt}21861}}


\bibitem{Connolly1983}
M.~Connolly.
\newblock {Analytical molecular surface calculation}.
\newblock {\em Journal of Applied Crystallography}, 16(5):548--558, Oct 1983.
  doi: {{%
10\hspace{.1pt}\discretionary{.}{%
}{.}\hspace{.4pt}1107\discretionary{/}{%
}{/}S0021889883010985}}


\bibitem{Durrant2014}
J.~D. Durrant and R.~E. Amaro.
\newblock {LipidWrapper}: An algorithm for generating large-scale membrane
  models of arbitrary geometry.
\newblock {\em PLOS Computational Biology}, 10(7):1--11, 2014. doi: {{%
10\hspace{.1pt}\discretionary{.}{%
}{.}\hspace{.4pt}1371\discretionary{/}{%
}{/}journal\hspace{.1pt}\discretionary{.}{%
}{.}\hspace{.4pt}pcbi\hspace{.1pt}\discretionary{.}{%
}{.}\hspace{.4pt}1003720}}


\bibitem{Ebert2002}
D.~S. Ebert, F.~K. Musgrave, D.~Peachey, K.~Perlin, and S.~Worley.
\newblock {\em Texturing and Modeling: A Procedural Approach}.
\newblock Morgan Kaufmann Publishers Inc., 3rd ed., 2002.

\bibitem{Edelsbrunner1994}
H.~Edelsbrunner and E.~P. M\"{u}cke.
\newblock Three-dimensional alpha shapes.
\newblock {\em ACM Transactions on Graphics}, 13(1):43--–72, 1994. doi: {{%
10\hspace{.1pt}\discretionary{.}{%
}{.}\hspace{.4pt}1145\discretionary{/}{%
}{/}174462\hspace{.1pt}\discretionary{.}{%
}{.}\hspace{.4pt}156635}}


\bibitem{Emsley2010}
P.~Emsley, B.~Lohkamp, W.~Scott, and K.~Cowtan.
\newblock Features and development of coot.
\newblock {\em Acta crystallographica. Section D, Biological crystallography},
  66:486--501, 04 2010. doi: {{%
10\hspace{.1pt}\discretionary{.}{%
}{.}\hspace{.4pt}1107\discretionary{/}{%
}{/}S0907444910007493}}


\bibitem{Falk2013}
M.~Falk, M.~Krone, and T.~Ertl.
\newblock Atomistic visualization of mesoscopic whole-cell simulations using
  ray-casted instancing.
\newblock {\em Computer Graphics Forum}, 32(8):195--206, 2013. doi: {{%
10\hspace{.1pt}\discretionary{.}{%
}{.}\hspace{.4pt}1111\discretionary{/}{%
}{/}cgf\hspace{.1pt}\discretionary{.}{%
}{.}\hspace{.4pt}12197}}


\bibitem{Galin2010}
E.~Galin, A.~Peytavie, N.~Maréchal, and E.~Guérin.
\newblock Procedural generation of roads.
\newblock {\em Computer Graphics Forum}, 29:429--438, 06 2010. doi: {{%
10\hspace{.1pt}\discretionary{.}{%
}{.}\hspace{.4pt}1111\discretionary{/}{%
}{/}j\hspace{.1pt}\discretionary{.}{%
}{.}\hspace{.4pt}1467\discretionary{%
}{-}{-}8659\hspace{.1pt}\discretionary{.}{%
}{.}\hspace{.4pt}2009\hspace{.1pt}\discretionary{.}{%
}{.}\hspace{.4pt}01612\hspace{.1pt}\discretionary{.}{%
}{.}\hspace{.4pt}x}}


\bibitem{Gardner2018}
A.~Gardner, L.~Autin, B.~Barbaro, A.~J. Olson, and D.~S. Goodsell.
\newblock Cellpaint: Interactive illustration of dynamic mesoscale cellular
  environments.
\newblock {\em IEEE Computer Graphics and Applications}, 38(6):51--66, 2018.
  doi: {{%
10\hspace{.1pt}\discretionary{.}{%
}{.}\hspace{.4pt}1109\discretionary{/}{%
}{/}MCG\hspace{.1pt}\discretionary{.}{%
}{.}\hspace{.4pt}2018\hspace{.1pt}\discretionary{.}{%
}{.}\hspace{.4pt}2877076}}


\bibitem{Goodsell2020}
D.~S. Goodsell, A.~J. Olson, and S.~Forli.
\newblock Art and science of the cellular mesoscale.
\newblock {\em Trends in Biochemical Sciences}, 2020. doi: {{%
10\hspace{.1pt}\discretionary{.}{%
}{.}\hspace{.4pt}1016\discretionary{/}{%
}{/}j\hspace{.1pt}\discretionary{.}{%
}{.}\hspace{.4pt}tibs\hspace{.1pt}\discretionary{.}{%
}{.}\hspace{.4pt}2020\hspace{.1pt}\discretionary{.}{%
}{.}\hspace{.4pt}02\hspace{.1pt}\discretionary{.}{%
}{.}\hspace{.4pt}010}}


\bibitem{Guerin2017}
E.~Gu\'{e}rin, J.~Digne, E.~Galin, A.~Peytavie, C.~Wolf, B.~Benes, and
  B.~Martinez.
\newblock Interactive example-based terrain authoring with conditional
  generative adversarial networks.
\newblock {\em ACM Transactions on Graphics}, 36(6), 2017. doi: {{%
10\hspace{.1pt}\discretionary{.}{%
}{.}\hspace{.4pt}1145\discretionary{/}{%
}{/}3130800\hspace{.1pt}\discretionary{.}{%
}{.}\hspace{.4pt}3130804}}


\bibitem{Gui2017}
M.~Gui, X.~Liu, D.~Guo, Z.~Zhang, C.-C. Yin, Y.~Chen, and Y.~Xiang.
\newblock Electron microscopy studies of the coronavirus ribonucleoprotein
  complex.
\newblock {\em Protein \& Cell}, 8(3):219--224, 2017. doi: {{%
10\hspace{.1pt}\discretionary{.}{%
}{.}\hspace{.4pt}1007\discretionary{/}{%
}{/}s13238\discretionary{%
}{-}{-}016\discretionary{%
}{-}{-}0352\discretionary{%
}{-}{-}8}}


\bibitem{Halladjian2020}
S.~Halladjian, H.~Miao, D.~Kou\v{r}il, M.~E. Gr\"{o}ller, I.~Viola, and
  T.~Isenberg.
\newblock {Scale Trotter}: Illustrative visual travels across negative scales.
\newblock {\em IEEE Transactions on Visualization and Computer Graphics},
  26(1):654--664, 2020.

\bibitem{Hermosilla2017}
P.~Hermosilla, M.~Krone, V.~Guallar, P.-P. Vázquez, A.~Vinacua, and
  T.~Ropinski.
\newblock Interactive gpu-based generation of solvent-excluded surfaces.
\newblock {\em The Visual Computer}, 33:869--–881, 2017. doi: {{%
doi\hspace{.1pt}\discretionary{.}{%
}{.}\hspace{.4pt}org\discretionary{/}{%
}{/}10\hspace{.1pt}\discretionary{.}{%
}{.}\hspace{.4pt}1007\discretionary{/}{%
}{/}s00371\discretionary{%
}{-}{-}017\discretionary{%
}{-}{-}1397\discretionary{%
}{-}{-}2}}


\bibitem{Igarashi1999}
T.~Igarashi, S.~Matsuoka, and H.~Tanaka.
\newblock Teddy: A sketching interface for 3d freeform design.
\newblock In {\em Proceedings of ACM SIGGRAPH ’99}, pp. 409--416, 1999. doi:
  {{%
10\hspace{.1pt}\discretionary{.}{%
}{.}\hspace{.4pt}1145\discretionary{/}{%
}{/}311535\hspace{.1pt}\discretionary{.}{%
}{.}\hspace{.4pt}311602}}


\bibitem{Jiang2020}
H.~{Jiang}, D.~{Yan}, X.~{Zhang}, and P.~{Wonka}.
\newblock Selection expressions for procedural modeling.
\newblock {\em IEEE Transactions on Visualization and Computer Graphics},
  26(4):1775--1788, 2020.

\bibitem{Johnson2014}
G.~Johnson, L.~Autin, M.~Al-Alusi, D.~Goodsell, M.~Sanner, and A.~Olson.
\newblock cellpack: A virtual mesoscope to model and visualize structural
  systems biology.
\newblock {\em Nature methods}, 12, 12 2014. doi: {{%
10\hspace{.1pt}\discretionary{.}{%
}{.}\hspace{.4pt}1038\discretionary{/}{%
}{/}nmeth\hspace{.1pt}\discretionary{.}{%
}{.}\hspace{.4pt}3204}}


\bibitem{Jumper2020}
J.~Jumper, K.~Tunyasuvunakool, P.~Kohli, D.~Hassabis, and the AlphaFold~Team.
\newblock {\em {Computational predictions of protein structures associated with
  COVID-19, version 2, DeepMind website, 8 April 2020}}.

\bibitem{Chang2014}
C.~ke~Chang, M.-H. Hou, C.-F. Chang, C.-D. Hsiao, and T.~huang Huang.
\newblock The sars coronavirus nucleocapsid protein – forms and functions.
\newblock {\em Antiviral Research}, 103:39--50, 2014. doi: {{%
10\hspace{.1pt}\discretionary{.}{%
}{.}\hspace{.4pt}1016\discretionary{/}{%
}{/}j\hspace{.1pt}\discretionary{.}{%
}{.}\hspace{.4pt}antiviral\hspace{.1pt}\discretionary{.}{%
}{.}\hspace{.4pt}2013\hspace{.1pt}\discretionary{.}{%
}{.}\hspace{.4pt}12\hspace{.1pt}\discretionary{.}{%
}{.}\hspace{.4pt}009}}


\bibitem{Kim2004}
D.~E. Kim, D.~Chivian, and D.~Baker.
\newblock Protein structure prediction and analysis using the {Robetta} server.
\newblock {\em Nucleic Acids Research}, 32(suppl\_2):W526--W531, 07 2004. doi:
  {{%
10\hspace{.1pt}\discretionary{.}{%
}{.}\hspace{.4pt}1093\discretionary{/}{%
}{/}nar\discretionary{/}{%
}{/}gkh468}}


\bibitem{Klein2018}
T.~Klein, L.~Autin, B.~Kozl\'{i}kov\'{a}, D.~S. Goodsell, A.~Olson, M.~E.
  Gr\"{o}ller, and I.~Viola.
\newblock Instant construction and visualization of crowded biological
  environments.
\newblock {\em IEEE Transactions on Visualization and Computer Graphics},
  24(1):862--872, 2018. doi: {{%
10\hspace{.1pt}\discretionary{.}{%
}{.}\hspace{.4pt}1109\discretionary{/}{%
}{/}TVCG\hspace{.1pt}\discretionary{.}{%
}{.}\hspace{.4pt}2017\hspace{.1pt}\discretionary{.}{%
}{.}\hspace{.4pt}2744258}}


\bibitem{Klein2019}
T.~Klein, P.~Mindek, L.~Autin, D.~Goodsell, A.~Olson, M.~E. Gr\"{o}ller, and
  I.~Viola.
\newblock Parallel generation and visualization of bacterial genome structures.
\newblock In {\em Computer Graphics Forum}, vol.~38, pp. 57--68, 2019. doi: {{%
10\hspace{.1pt}\discretionary{.}{%
}{.}\hspace{.4pt}1111\discretionary{/}{%
}{/}cgf\hspace{.1pt}\discretionary{.}{%
}{.}\hspace{.4pt}13816}}


\bibitem{Krone2009}
M.~Krone, K.~Bidmon, and T.~Ertl.
\newblock Interactive visualization of molecular surface dynamics.
\newblock {\em IEEE Transactions on Visualization and Computer Graphics},
  15(6):1391--1398, 2009.

\bibitem{LeMuzic2014}
M.~Le~Muzic, J.~Parulek, A.-K. Stavrum, and I.~Viola.
\newblock Illustrative visualization of molecular reactions using omniscient
  intelligence and passive agents.
\newblock {\em EG Computer Graphics Forum}, 33(3):141--150, 2014. doi: {{%
10\hspace{.1pt}\discretionary{.}{%
}{.}\hspace{.4pt}1111\discretionary{/}{%
}{/}cgf\hspace{.1pt}\discretionary{.}{%
}{.}\hspace{.4pt}12370}}


\bibitem{Lidal2013}
E.~M. Lidal, M.~Natali, D.~Patel, H.~Hauser, and I.~Viola.
\newblock Geological storytelling.
\newblock {\em Computers \& Graphics}, 37(5):445--459, 2013. doi: {{%
10\hspace{.1pt}\discretionary{.}{%
}{.}\hspace{.4pt}1016\discretionary{/}{%
}{/}j\hspace{.1pt}\discretionary{.}{%
}{.}\hspace{.4pt}cag\hspace{.1pt}\discretionary{.}{%
}{.}\hspace{.4pt}2013\hspace{.1pt}\discretionary{.}{%
}{.}\hspace{.4pt}01\hspace{.1pt}\discretionary{.}{%
}{.}\hspace{.4pt}010}}


\bibitem{Lindenmayer1968}
A.~Lindenmayer.
\newblock Mathematical models for cellular interactions in development i.
  filaments with one-sided inputs.
\newblock {\em Journal of Theoretical Biology}, 18(3):280--299, 1968. doi: {{%
10\hspace{.1pt}\discretionary{.}{%
}{.}\hspace{.4pt}1016\discretionary{/}{%
}{/}0022\discretionary{%
}{-}{-}5193\discretionary{%
}{(}{(}68\discretionary{)}{%
}{)}90079\discretionary{%
}{-}{-}9}}


\bibitem{Lindow2012}
N.~Lindow, D.~Baum, and H.-C. Hege.
\newblock Interactive rendering of materials and biological structures on
  atomic and nanoscopic scale.
\newblock {\em Computer Graphics Forum}, 31(3pt4):1325--1334, 2012. doi: {{%
10\hspace{.1pt}\discretionary{.}{%
}{.}\hspace{.4pt}1111\discretionary{/}{%
}{/}j\hspace{.1pt}\discretionary{.}{%
}{.}\hspace{.4pt}1467\discretionary{%
}{-}{-}8659\hspace{.1pt}\discretionary{.}{%
}{.}\hspace{.4pt}2012\hspace{.1pt}\discretionary{.}{%
}{.}\hspace{.4pt}03128\hspace{.1pt}\discretionary{.}{%
}{.}\hspace{.4pt}x}}


\bibitem{Liu2020}
C.~Liu, Y.~Yang, Y.~Gao, C.~Shen, B.~Ju, C.~Liu, X.~Tang, J.~Wei, X.~Ma,
  W.~Liu, S.~Xu, Y.~Liu, J.~Yuan, J.~Wu, Z.~Liu, Z.~Zhang, P.~Wang, and L.~Liu.
\newblock Viral architecture of sars-cov-2 with post-fusion spike revealed by
  cryo-em.
\newblock {\em bioRxiv}, 2020. doi: {{%
10\hspace{.1pt}\discretionary{.}{%
}{.}\hspace{.4pt}1101\discretionary{/}{%
}{/}2020\hspace{.1pt}\discretionary{.}{%
}{.}\hspace{.4pt}03\hspace{.1pt}\discretionary{.}{%
}{.}\hspace{.4pt}02\hspace{.1pt}\discretionary{.}{%
}{.}\hspace{.4pt}972927}}


\bibitem{Mcneughton1978}
M.~R. Macnaughton, H.~A. Davies, and M.~V. Nermut.
\newblock Ribonucleoprotein-like structures from coronavirus particles.
\newblock {\em Journal of General Virology}, 39(3):545--549, 1978. doi: {{%
10\hspace{.1pt}\discretionary{.}{%
}{.}\hspace{.4pt}1099\discretionary{/}{%
}{/}0022\discretionary{%
}{-}{-}1317\discretionary{%
}{-}{-}39\discretionary{%
}{-}{-}3\discretionary{%
}{-}{-}545}}


\bibitem{Mindek2018}
P.~Mindek, D.~Kou\v{r}il, J.~Sorger, D.~Toloudis, B.~Lyons, G.~Johnson, M.~E.
  Gr{\"o}ller, and I.~Viola.
\newblock Visualization multi-pipeline for communicating biology.
\newblock {\em IEEE Transactions on Visualization and Computer Graphics},
  24(1):883--892, 2018. doi: {{%
10\hspace{.1pt}\discretionary{.}{%
}{.}\hspace{.4pt}1109\discretionary{/}{%
}{/}TVCG\hspace{.1pt}\discretionary{.}{%
}{.}\hspace{.4pt}2017\hspace{.1pt}\discretionary{.}{%
}{.}\hspace{.4pt}2744518}}


\bibitem{Mueller2006}
P.~M\"{u}ller, P.~Wonka, S.~Haegler, A.~Ulmer, and L.~Van~Gool.
\newblock Procedural modeling of buildings.
\newblock {\em ACM Transactions on Graphics}, 25(3):614--623, 2006. doi: {{%
10\hspace{.1pt}\discretionary{.}{%
}{.}\hspace{.4pt}1145\discretionary{/}{%
}{/}1141911\hspace{.1pt}\discretionary{.}{%
}{.}\hspace{.4pt}1141931}}


\bibitem{LeMuzic2015}
M.~L. Muzic, L.~Autin, J.~Parulek, and I.~Viola.
\newblock {cellVIEW}: a tool for illustrative and multi-scale rendering of
  large biomolecular datasets.
\newblock In {\em Proceedings of EG VBCM}, pp. 61--70, 2015. doi: {{%
10\hspace{.1pt}\discretionary{.}{%
}{.}\hspace{.4pt}2312\discretionary{/}{%
}{/}vcbm\hspace{.1pt}\discretionary{.}{%
}{.}\hspace{.4pt}20151209}}


\bibitem{Natali2014}
M.~Natali, J.~Parulek, and D.~Patel.
\newblock Rapid modelling of interactive geological illustrations with faults
  and compaction.
\newblock In {\em Proceedings of the 30th Spring Conference on Computer
  Graphics}, pp. 5--12, 2014. doi: {{%
10\hspace{.1pt}\discretionary{.}{%
}{.}\hspace{.4pt}1145\discretionary{/}{%
}{/}2643188\hspace{.1pt}\discretionary{.}{%
}{.}\hspace{.4pt}2643201}}


\bibitem{Neuman2016}
B.~Neuman and M.~Buchmeier.
\newblock Chapter one - supramolecular architecture of the coronavirus
  particle.
\newblock In {\em Coronaviruses}, vol.~96, pp. 1--27. Academic Press, 2016.
  doi: {{%
10\hspace{.1pt}\discretionary{.}{%
}{.}\hspace{.4pt}1016\discretionary{/}{%
}{/}bs\hspace{.1pt}\discretionary{.}{%
}{.}\hspace{.4pt}aivir\hspace{.1pt}\discretionary{.}{%
}{.}\hspace{.4pt}2016\hspace{.1pt}\discretionary{.}{%
}{.}\hspace{.4pt}08\hspace{.1pt}\discretionary{.}{%
}{.}\hspace{.4pt}005}}


\bibitem{Neuman2006}
B.~W. Neuman, B.~D. Adair, C.~Yoshioka, J.~D. Quispe, G.~Orca, P.~Kuhn, R.~A.
  Milligan, M.~Yeager, and M.~J. Buchmeier.
\newblock Supramolecular architecture of severe acute respiratory syndrome
  coronavirus revealed by electron cryomicroscopy.
\newblock {\em Journal of Virology}, 80(16):7918--7928, 2006. doi: {{%
10\hspace{.1pt}\discretionary{.}{%
}{.}\hspace{.4pt}1128\discretionary{/}{%
}{/}JVI\hspace{.1pt}\discretionary{.}{%
}{.}\hspace{.4pt}00645\discretionary{%
}{-}{-}06}}


\bibitem{Nishida2016}
G.~Nishida, I.~Garcia-Dorado, D.~G. Aliaga, B.~Benes, and A.~Bousseau.
\newblock Interactive sketching of urban procedural models.
\newblock {\em ACM Transactions on Graphics}, 35(4), 2016. doi: {{%
10\hspace{.1pt}\discretionary{.}{%
}{.}\hspace{.4pt}1145\discretionary{/}{%
}{/}2897824\hspace{.1pt}\discretionary{.}{%
}{.}\hspace{.4pt}2925951}}


\bibitem{ODonoghue2010}
S.~O'Donoghue, A.-C. Gavin, N.~Gehlenborg, D.~S. Goodsell, J.-K. Hériché,
  C.~Nielsen, C.~North, A.~Olson, J.~Procter, D.~Shattuck, T.~Walter, and
  B.~Wong.
\newblock Visualizing biological data—now and in the future.
\newblock {\em Nature Methods}, 7(3):S2, 2010. doi: {{%
10\hspace{.1pt}\discretionary{.}{%
}{.}\hspace{.4pt}1038\discretionary{/}{%
}{/}nmeth\hspace{.1pt}\discretionary{.}{%
}{.}\hspace{.4pt}f\hspace{.1pt}\discretionary{.}{%
}{.}\hspace{.4pt}301}}


\bibitem{Parish2001}
Y.~I.~H. Parish and P.~M\"{u}ller.
\newblock Procedural modeling of cities.
\newblock In {\em Proceedings of the 28th Annual Conference on Computer
  Graphics and Interactive Techniques}, pp. 301--308. Association for Computing
  Machinery, 2001. doi: {{%
10\hspace{.1pt}\discretionary{.}{%
}{.}\hspace{.4pt}1145\discretionary{/}{%
}{/}383259\hspace{.1pt}\discretionary{.}{%
}{.}\hspace{.4pt}383292}}


\bibitem{Parulek2013}
J.~Parulek and A.~Brambilla.
\newblock Fast blending scheme for molecular surface representation.
\newblock {\em IEEE Transactions on Visualization and Computer Graphics},
  19(12):2653--2662, 2013.

\bibitem{Pirhadi2016}
S.~Pirhadi, J.~Sunseri, and D.~R. Koes.
\newblock Open source molecular modeling.
\newblock {\em Journal of Molecular Graphics and Modelling}, 69:127--143, 2016.
  doi: {{%
10\hspace{.1pt}\discretionary{.}{%
}{.}\hspace{.4pt}1016\discretionary{/}{%
}{/}j\hspace{.1pt}\discretionary{.}{%
}{.}\hspace{.4pt}jmgm\hspace{.1pt}\discretionary{.}{%
}{.}\hspace{.4pt}2016\hspace{.1pt}\discretionary{.}{%
}{.}\hspace{.4pt}07\hspace{.1pt}\discretionary{.}{%
}{.}\hspace{.4pt}008}}


\bibitem{Portenier2018}
T.~Portenier, Q.~Hu, A.~Szab\'{o}, S.~A. Bigdeli, P.~Favaro, and M.~Zwicker.
\newblock Faceshop: Deep sketch-based face image editing.
\newblock {\em ACM Transactions on Graphics}, 37(4), 2018. doi: {{%
10\hspace{.1pt}\discretionary{.}{%
}{.}\hspace{.4pt}1145\discretionary{/}{%
}{/}3197517\hspace{.1pt}\discretionary{.}{%
}{.}\hspace{.4pt}3201393}}


\bibitem{Prusinkiewicz1990}
P.~Prusinkiewicz and A.~Lindenmayer.
\newblock {\em The algorithmic beauty of plants}.
\newblock Springer-Verlag New York, Inc., 1990.

\bibitem{Richards1977}
F.~M. Richards.
\newblock Areas, volumes, packing, and protein structure.
\newblock {\em Annual Review of Biophysics and Bioengineering}, 6(1):151--176,
  1977. doi: {{%
10\hspace{.1pt}\discretionary{.}{%
}{.}\hspace{.4pt}1146\discretionary{/}{%
}{/}annurev\hspace{.1pt}\discretionary{.}{%
}{.}\hspace{.4pt}bb\hspace{.1pt}\discretionary{.}{%
}{.}\hspace{.4pt}06\hspace{.1pt}\discretionary{.}{%
}{.}\hspace{.4pt}060177\hspace{.1pt}\discretionary{.}{%
}{.}\hspace{.4pt}001055}}


\bibitem{Sanner1996}
M.~F. Sanner, A.~J. Olson, and J.-C. Spehner.
\newblock Reduced surface: An efficient way to compute molecular surfaces.
\newblock {\em Biopolymers}, 38(3):305--320, 1996. doi: {{%
10\hspace{.1pt}\discretionary{.}{%
}{.}\hspace{.4pt}1002\discretionary{/}{%
}{/}\discretionary{%
}{(}{(}SICI\discretionary{)}{%
}{)}1097\discretionary{%
}{-}{-}0282\discretionary{%
}{(}{(}199603\discretionary{)}{%
}{)}38\discretionary{:}{%
}{:}3{\textless}305\discretionary{:}{%
}{:}\discretionary{:}{%
}{:}AID\discretionary{%
}{-}{-}BIP4{\textgreater}3\hspace{.1pt}\discretionary{.}{%
}{.}\hspace{.4pt}0\hspace{.1pt}\discretionary{.}{%
}{.}\hspace{.4pt}CO\discretionary{;}{%
}{;}2\discretionary{%
}{-}{-}Y}}


\bibitem{Schatz2016}
K.~Schatz, C.~M\"{u}ller, M.~Krone, J.~Schneider, G.~Reina, and T.~Ertl.
\newblock Interactive visual exploration of a trillion particles.
\newblock In IEEE, ed., {\em Symposium on Large Data Analysis and Visualization
  (LDAV)}, 2016. doi: {{%
10\hspace{.1pt}\discretionary{.}{%
}{.}\hspace{.4pt}1109\discretionary{/}{%
}{/}LDAV\hspace{.1pt}\discretionary{.}{%
}{.}\hspace{.4pt}2016\hspace{.1pt}\discretionary{.}{%
}{.}\hspace{.4pt}7874310}}


\bibitem{PyMOL2015}
L.~Schr\"odinger.
\newblock The {PyMOL} molecular graphics system, version~1.8.
\newblock 2015.

\bibitem{Schwarz2015}
M.~Schwarz and P.~M{\"u}ller.
\newblock Advanced procedural modeling of architecture.
\newblock {\em ACM Transactions on Graphics}, 34(4):107:1--107:12, 2015.

\bibitem{Singh2019}
A.~Singh, D.~Montgomery, X.~Xue, B.~L. Foley, and R.~J. Woods.
\newblock Gag builder: a web-tool for modeling 3d structures of
  glycosaminoglycans.
\newblock {\em Glycobiology}, 29(7):515--518, 04 2019. doi: {{%
10\hspace{.1pt}\discretionary{.}{%
}{.}\hspace{.4pt}1093\discretionary{/}{%
}{/}glycob\discretionary{/}{%
}{/}cwz027}}


\bibitem{Tarini2006}
M.~Tarini, P.~Cignoni, and C.~Montani.
\newblock Ambient occlusion and edge cueing for enhancing real time molecular
  visualization.
\newblock {\em IEEE Transactions on Visualization and Computer Graphics},
  12(5):1237--1244, 2006. doi: {{%
10\hspace{.1pt}\discretionary{.}{%
}{.}\hspace{.4pt}1109\discretionary{/}{%
}{/}TVCG\hspace{.1pt}\discretionary{.}{%
}{.}\hspace{.4pt}2006\hspace{.1pt}\discretionary{.}{%
}{.}\hspace{.4pt}115}}


\bibitem{Torpa2006}
J.~Torppa, J.~P.~T. Valkonen, and K.~Muinonen.
\newblock Three-dimensional stochastic shape modelling for potato tubers.
\newblock {\em Potato Research}, 49(2):109--118, 2006. doi: {{%
10\hspace{.1pt}\discretionary{.}{%
}{.}\hspace{.4pt}1007\discretionary{/}{%
}{/}s11540\discretionary{%
}{-}{-}006\discretionary{%
}{-}{-}9010\discretionary{%
}{-}{-}5}}


\bibitem{Varshney1994}
A.~Varshney, F.~P. {Brooks, Jr.}, and W.~V. Wright.
\newblock Linearly scalable computation of smooth molecular surfaces.
\newblock {\em IEEE Computer Graphics and Applications}, 14(5):19--25, 1994.

\bibitem{Waterhouse2018}
A.~Waterhouse, M.~Bertoni, S.~Bienert, G.~Studer, G.~Tauriello, R.~Gumienny,
  F.~T. Heer, T.~A. de Beer, C.~Rempfer, L.~Bordoli, R.~Lepore, and
  T.~Schwede.
\newblock Swiss-model: homology modelling of protein structures and complexes.
\newblock {\em Nucleic Acids Research}, 46(W1):W296--W303, 05 2018. doi: {{%
10\hspace{.1pt}\discretionary{.}{%
}{.}\hspace{.4pt}1093\discretionary{/}{%
}{/}nar\discretionary{/}{%
}{/}gky427}}


\bibitem{Webanck2018}
A.~Webanck, Y.~Cortial, E.~Guérin, and E.~Galin.
\newblock Procedural cloudscapes.
\newblock {\em Computer Graphics Forum}, 37(2):431--442, 2018. doi: {{%
10\hspace{.1pt}\discretionary{.}{%
}{.}\hspace{.4pt}1111\discretionary{/}{%
}{/}cgf\hspace{.1pt}\discretionary{.}{%
}{.}\hspace{.4pt}13373}}


\bibitem{Williams1991}
L.~Williams.
\newblock Shading in two dimensions.
\newblock In {\em Proceedings of Graphics Interface}, pp. 143--151, 1991. doi:
  {{%
10\hspace{.1pt}\discretionary{.}{%
}{.}\hspace{.4pt}20380\discretionary{/}{%
}{/}GI1991\hspace{.1pt}\discretionary{.}{%
}{.}\hspace{.4pt}19}}


\bibitem{Wu2020}
F.~Wu, S.~Zhao, B.~Yu, Y.-M. Chen, W.~Wang, Z.-G. Song, Y.~Hu, Z.-W. Tao, J.-H.
  Tian, Y.-Y. Pei, et~al.
\newblock A new coronavirus associated with human respiratory disease in china.
\newblock {\em Nature}, 579(7798):265--269, 2020.

\end{thebibliography}

\appendix

\clearpage
\centering
\section*{Appendix}
\SetAlgorithmName{Algorithm}{algorithm}{List of Algorithms}
\begin{algorithm}
\SetAlgoLined
  \SetKwInOut{Input}{Input}
  \SetKwInOut{Output}{Output}
  \SetKwProg{GenerateContour}{GenerateContour}{}{}
 
 \GenerateContour{$(C, N_{sample})$}{
    \Input{Contours set $C$; \# angle directions $N_{sample}$}
    \Output{Generated contour $C_{g}$} 

   //Augmented data\\
    $C_{stat} = []$\; 
    \ForEach{$c \in C$}{
        $cp$ = $c$.Rotate($\pi$)\;
        $cf$ = $c$.Flip(x-axis)\;
        $cpf$ = $cp$.Flip(x-axis)\;
        $C_{stat}$.add($c, cp, cf, cpf$)\;
    }
    
    //Get max distance from all points on contours to origin\\
    $R_{stat} = []$\; 
    $angle_{sample} = []$\;    
    \ForEach{$c \in C_{stat}$}{
        \ForEach{point $p \in c$}{
            $R_{stat}$.add($Distance(p, origin)$)\;
        }
    }
    $r_{max} = Max(R_{stat})$\;
    
    //Cast a ray from origin to all directions\\
    $i = 0$\;
    \While{$i++ < N_{sample}$}{ 
        $angle_{sample}$.add($2i\pi/N_{sample}$)\;
    }

    $Ray_{sample} = array(N_{sample})$\; 
    $i = 0$\;
    \While{$i++ < N_{sample}$}{
        $Ray_{sample}[i].x$ = $r_{max} cos(angle_{sample}[i])$\;
        $Ray_{sample}[i].y$ = $r_{max} sin(angle_{sample}[i])$\;
    }
    // Fit normal distribution of $r_{\cap}$ for each angle

    \ForEach{$r \in Ray_{sample}$}{

        $intersections = []$\;
        \ForEach{$c \in C_{stat}$}{
            $intersections$.add($Intersect(r, c)$)\;
        }
        $r_{\cap} = Distance(Intersect(r, c), origin)$\;
        $\sigma_{s}$ = StandardDeviation($r_{\cap}$)\;
        $\mu_{s}$ = Mean($r_{\cap}$)\;
        $min_{s}$ = Min($r_{\cap}$)\;
        $max_{s}$ = Max($r_{\cap}$)\;
    }
    
    //Rejection sampling for each angle\\
    $C_{g} = []\;$
    \ForEach{$\theta_{s} \in angle_{sample}$}{
        $C_g.add(RejectionSampling(\theta_{s}, min_{s}, max_{s}, \mu_{s}, \sigma_{s}))$\;
    }
    
    //Interpolate $C_{g}$\\
    $C_{g} = CatmullRom(C_{g})$\; 
 }
\caption{Generate new contour from extracted contours.}
\label{alg:generate-contours}
\end{algorithm}

\SetAlgorithmName{Algorithm}{algorithm}{List of Algorithms}
\begin{algorithm}
\SetAlgoLined
  \SetKwInOut{Input}{Input}
  \SetKwInOut{Output}{Output}
  \SetKwProg{RejectionSampling}{RejectionSampling}{}{}
 
 \RejectionSampling{$(\theta_{s}, min_{s}, max_{s}, \mu_{s}, \sigma_{s})$}{
    \Input{Angle $\theta_{s}$\; distribution of $r_{\cap}$ of $\theta_{s}$: $min_{s}, max_{s}, \mu_{s}, \sigma_{s}$}
    \Output{One sample point of new contour $p_{g}$} 
    
    $accept = false$\;
    \While {!$accept$}{
        Choose $r \sim U(min_{s}, max_{s})$\;
        Choose $prob \sim U(0, MaxProbabilityDensity(r_{\cap}))$\; 
        $pdf(r) = ProbabilityDensity(r)$\;
    
        \If{$pdf(r) \leq prob$}{
            $accept = true$\;
            $p_{g}.x$ = $rcos(\theta_{s})$\;
            $p_{g}.y$ = $rsin(\theta_{s})$\;
        }
    }
 }
\caption{Rejection sampling for create new contour.}
\label{alg:rejection-sampling}
\end{algorithm}
\setcounter{algocf}{1}

\SetAlgorithmName{Data structure}{data structure}{List of Data Structures}
\begin{algorithm}
  \SetAlgoNoLine\PrintSemicolon
  
  \Struct{Rule}{
    \String{} $Name_{in}$, $Name_{out}$; // identifier of elements $E(Name_{in})$, $E(Name_{out})$\\
    \Int{} $count$; // temporary counter of generated elements\\
    \Int{} $count_{max}$; // number of elements to be placed\\
    \Enum{} $distanceDistributionType$ // type of distribution function used [Uniform, Gaussian]\\
    \Int{} $collisions$; // temporary counter of detected collisions\\
    \Int{} $collisions_{max}$; // maximal number of errors while collisions computation\\
    \String{} $groupName$; // what group the rule belongs to\\
    \Float{} $probability$; // user-defined probability\\
    \Int{} $T_{num}$; // number of transformations processed\\
    \Float{} $yaw|pitch|roll$; // user-defined maximal allowed deviation in rotation\\     
    \Float{} $T_i.distance$; // distance of $E(Name_{in})$ to $E(Name_{out})$\\
    \Float{} $T_i.distanceStdDev$; // standard deviation of a generated distance (default=0)\\
    \Float{} $T_i.probability$; // probability of selecting the transformation\\
    \Quat{} $T_i.rotation$; // rotation of $E(Name_{out})$\\
    \Vect{} $T_i.translation$; // translation from $E(Name_{in})$ to $E(Name_{out})$\\
    \Quat{} $T_i.angularDifference$; // rotation from vector $(S, E(Name_{in}))$ to vector $(S, E(Name_{out}))$ if bound to a skeleton $S$\\
    \Int{} $T_i.v_j$; // closest vertex of the bound skeleton\\
  }
  
  \caption{Description of a rule $r$ data structure.} 
  \label{alg:rule-definition}
\end{algorithm}

\end{document}